\documentclass[a4paper,11pt]{article}
\usepackage{jinstpub} % for details on the use of the package, please see the JINST-author-manual
\DeclareMathOperator*{\argmin}{argmin}
\title{\boldmath Towards CRES-Based Non-destructive Electron Momentum Estimation for the PTOLEMY Relic Neutrino Detector}

\author[a,b]{Yuno Iwasaki,}
\author[a]{Andi Tan}
 \author[a,1 ]{and Christopher G. Tully\note{Corresponding author}}
 \affiliation[a]{Princeton University,\\Princeton NJ, USA}
 \affiliation[b]{University of California, Berkeley,\\Berkeley CA, USA}

\emailAdd{cgtully@princeton.edu}

\abstract{
The novel electron spectrometry method proposed by the PTOLEMY relic neutrino experiment requires a real-time, \textit{non-destructive} estimate of the parallel and transverse momentum splits of tritium $\beta$-decay electrons. The collaboration has proposed to obtain this estimate using cyclotron-radiation emission spectroscopy (CRES), in which the kinetic energy of a charged particle is determined by measuring the relativistic frequency shift of the cyclotron radiation emitted by the particle in a magnetic field. However, no suitable approach to extract this information in a non-destructive manner has been developed to date. In this paper, we characterize the performance of a configuration that can be feasibly integrated directly into the existing design for the transverse drift filter proposed by the PTOLEMY collaboration.
We study a geometry incorporating a cavity resonator to enhance a ${\sim}\mathcal{O}(1) \hspace{1mm}\mathrm{fW}$ cyclotron radiation signal and derive key features of the expected observed radiation specific to our radio-frequency~(RF) tracking configuration. We estimate the performance of our design using electromagnetic simulations and propose a general signal reconstruction algorithm capable of matching an observed signal to electron kinematic parameters. The projected signal-to-noise ratio~(SNR) of this technique suggests that a non-destructive RF tracking system based on an array of these components as building blocks is applicable for extracting the kinematic parameters of tritium endpoint electrons to the precision required for the PTOLEMY experiment.
}


\begin{document}
\maketitle
\flushbottom
\newpage
\section{Introduction}
\label{sec:intro}

Standard cosmological models generically predict a relic sea of low-energy neutrinos known as the cosmic neutrino background (C$\nu$B) \cite{Peebles1994}\cite{mukhanov}. C$\nu$B detection has the potential to uncover new physics in both cosmology and neutrino physics, but it has not yet been achieved due to the immense technical difficulty of detecting low-energy neutrinos \cite{mukhanov}\cite{ptolemy_2019}. Relic neutrinos are predicted to have momenta of order ${\sim} \mathcal{O}(10^{-4}) \hspace{1mm}\mathrm{eV}/c$ today,  requiring an interaction with effectively zero neutrino energy threshold and sufficiently large interaction cross section for vanishing neutrino momenta \cite{mukhanov}\cite{neutrino_cosmology}. This makes neutrino capture on $\beta$-decaying nuclei (NCB) a promising approach. First proposed by Steven Weinberg in 1962, then later formulated for massive neutrinos by Cocco, Mangano, and Messina in 2007, this method proposes to detect relic neutrino-induced inverse $\beta$-decay of an unstable nucleus that undergoes regular $\beta$-decay \cite{weinberg}\cite{cocco_mangano_messina}. The signature of relic neutrino capture is a narrow peak of electron counts at an energy $2m_{\beta}$ above the endpoint energy of the corresponding $\beta$-decay spectrum.  $m_{\beta} = \sqrt{\sum_{i=1}^3 |U_{ei}|^2 m_i^2}$ is the effective electron neutrino mass contributing kinematically to the $\beta$-decay, given by the incoherent sum of the masses of the neutrino mass eigenstates $\{m_i\}_{i = 1}^{3}$ and the electronic mixing matrix elements $U_{ei}$. The current best model-independent bound on this value is $m_{\beta}< 0.8\,\mathrm{eV}/c^2$ and was set by the KATRIN experiment by measuring the distortion of the tail of the tritium $\beta$-spectrum due to the finite neutrino mass~\cite{katrin_nature}. Additionally, a bound on the sum of the neutrino masses $\sum_{i=1}^3 m_i$ can be derived from the anisotropies of the CMB assuming a $\Lambda$CDM model; the current best constraint is $\sum_{i=1}^3 m_i < 0.12\,\mathrm{eV}/c^2$ and was determined by the Planck collaboration \cite{pdg}\cite{planck2020}. \\

\begin{figure}[htbp]
    \centering
    \includegraphics[width=0.95\columnwidth]{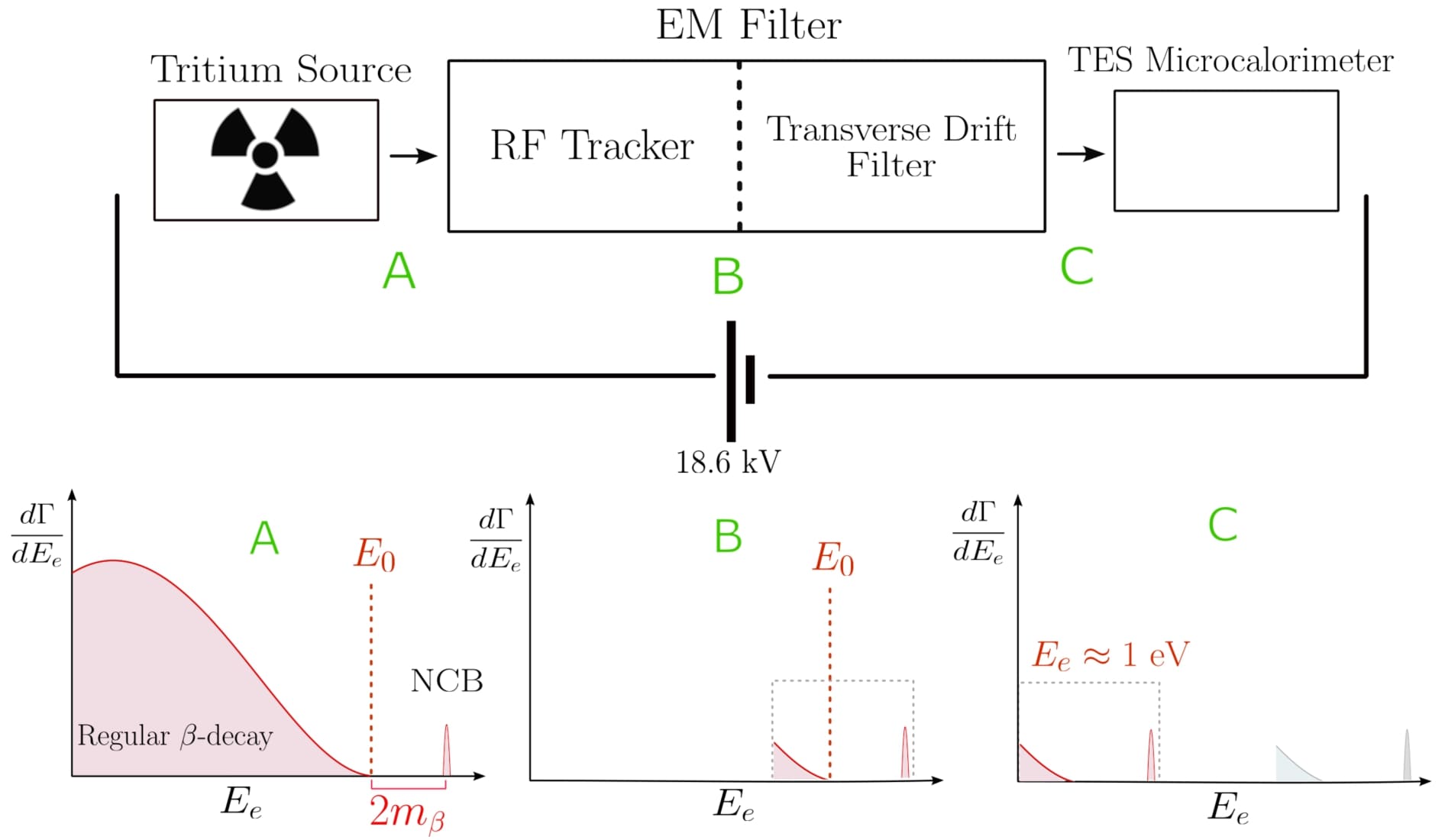}
    \caption{Schematic of the PTOLEMY experiment. A, B, and C represent the spectra of electrons reaching each experiment stage. Note that the NCB peak and regular $\beta$-decay spectra are not to scale; the NCB peak is much smaller. Furthermore, the two spectra are separated by twice $m_{\beta}$ (defined in the first paragraph of the introduction). A detailed analysis of the relative event rates due to regular $\beta$-decay and NCB can be found in \cite{ptolemy_2019}, Section 3.\label{fig:spectrum_schematic}}
\end{figure}

The PTOLEMY collaboration proposes to detect relic neutrinos by precision spectroscopy of the tritium $(\mathrm{H}^{3})$ $\beta$-decay spectrum near its $E_e {\approx} 18.6 \hspace{1mm}\mathrm{keV}$ endpoint energy to search for a small peak of counts due to inverse $\beta$-decay induced by relic neutrinos. Tritium was selected for its relatively low Q-value \cite{neutrino_cosmology}\cite{cocco_mangano_messina}\cite{nudat}, large NCB cross section \cite{neutrino_cosmology}\cite{cocco_mangano_messina}, and availability \cite{ptolemy_2019}. The basic principle for electron spectroscopy in the PTOLEMY experiment is the following. Using a configuration of static electromagnetic fields, $\beta$-decay electrons emitted by a solid-state tritium source \cite{ptolemy_2019} with energies in the vicinity of the ${\sim} 18.6\,\mathrm{keV}$ endpoint are guided to a calorimeter, while reducing their kinetic energy down to a value of order ${\sim} 1\,\mathrm{eV}$. Provided that the voltage difference between the source and calorimeter is kept stable (to the ppm level), the difference in the initial and final energy for each emitted electron is nearly the same, with the exception of variations in the final kinetic energy due to radiative energy losses. Thus, the shape of the electron spectrum near the endpoint energy will be reproduced at the calorimeter, but shifted to a lower energy and accompanied by some smearing due to RF emissions (before correction).  A schematic of this concept is provided in Figure~\ref{fig:spectrum_schematic}. Reducing electron kinetic energies is necessary to reach the dynamical range of the Transition Edge Sensor (TES) microcalorimeter, which will measure the electron spectrum with a final resolution of $50\,{\mathrm{meV}}$ \cite{ptolemy_2019}.\\

\begin{figure}[htbp]
    \centering
    \includegraphics[width=0.9\columnwidth]{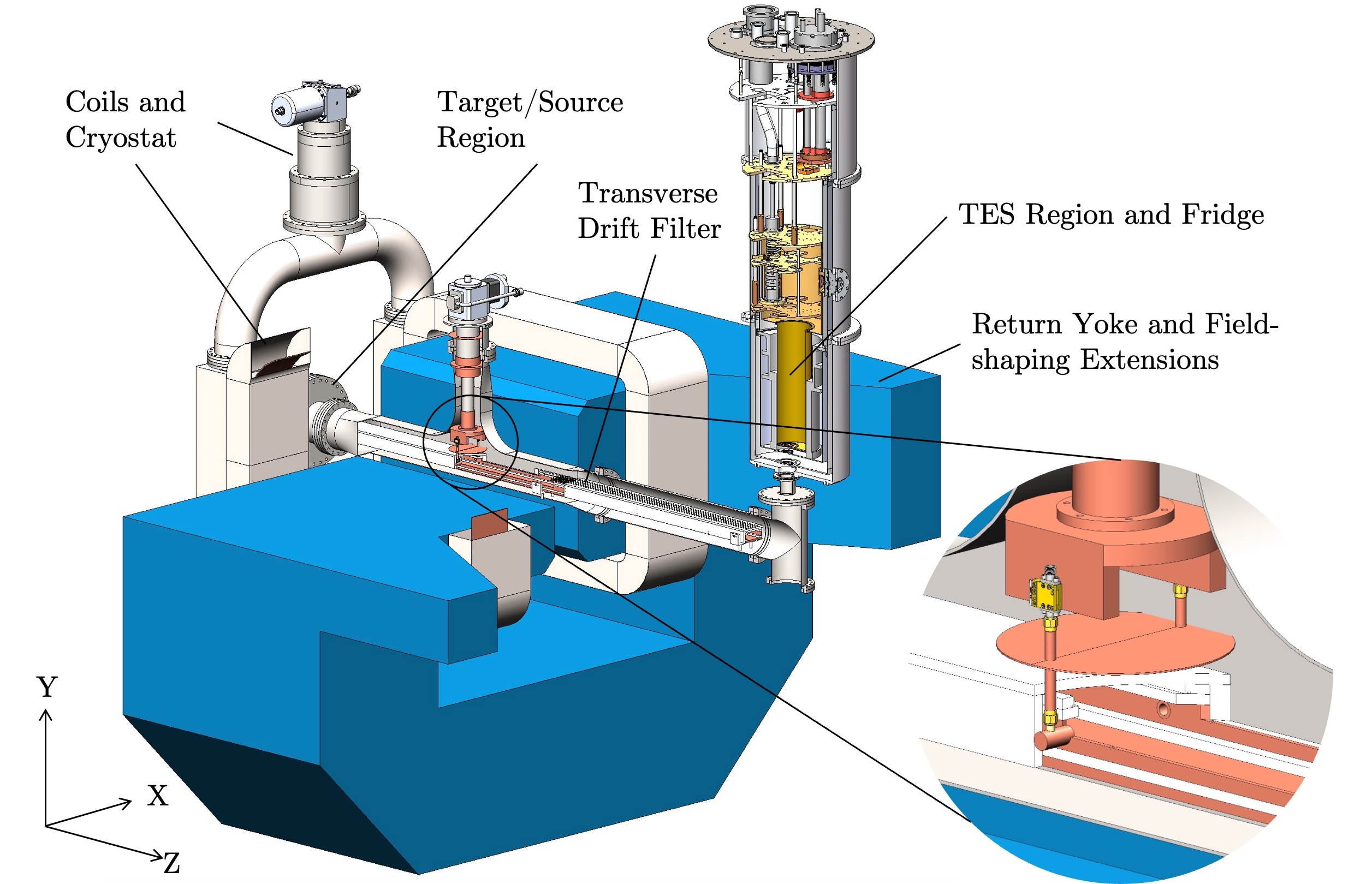}
    \caption{Schematic of the PTOLEMY demonstrator setup. Central to the design, the elongated rectangular structure along the $z$-axis is the EM filter, while the conductive cooling superconducting coils source the magnetic field. Within the highlighted section, RF tracking occurs in the uniform $B$-field region between the pole faces, while transverse drift occurs in the decaying $B$-field region between the fringe field shaping extensions. Electrode potentials within the RF tracking zone are static. Meanwhile, the filter electrodes in the transverse drift region are dynamically adjusted to produce the appropriate electric field for a given electron $K_{\perp}$. \label{fig:ptolemyconcept}}
\end{figure}

Electron kinetic energy reduction will be accomplished by a method invented by the PTOLEMY collaboration called the \textit{transverse drift filter}, which uses a configuration of static electric and magnetic fields to guide the electrons from the source to the calorimeter in a controlled motion referred to as \textit{transverse drift}. A detailed description of the transverse drift mechanism and the conceptual design of the filter is presented in \cite{EM_filter_design}; implementation and optimization of the transverse drift filter is presented in \cite{EM_filter_implementation}. Previous measurements of the tritium endpoint have been enabled by a technology called Magnetic Adiabatic Collimation and Electromagnetic (MAC-E)
filter, implemented by the Troitsk \cite{troitsk_final}, Mainz \cite{mainz_final}, and KATRIN neutrino mass experiments \cite{katrin_nature}.  The primary advantage of the transverse drift filter over previous techniques for electromagnetic filtering is its compact size: each dimension of the conceptual design shown in Figure~\ref{fig:ptolemyconcept} is ${\approx} 1\hspace{1mm}\mathrm{m}$. This feature is crucial for enabling the construction and simultaneous operation of multiple relic neutrino detectors, to realize the long-term goal of developing a C$\nu$B anisotropy map analogous to that of the cosmic microwave background \cite{ptolemy_2019}\cite{EM_filter_implementation}. However, a crucial trade-off of the exceptionally compact size of the filter is the requirement of an estimate of $K_{\perp}$---the portion of the electron kinetic energy due to its transverse momentum---to a precision of ${\sim} 10\hspace{1mm}\mathrm{eV}$, before the electron enters the transverse drift filter \cite{EM_filter_design}\cite{EM_filter_implementation}. Electron momentum split estimation must be performed \textit{non-destructively}, so that the trajectory of the electron from source to calorimeter is not obstructed.\\

%\begin{figure}[htbp]
%    \centering
%    \includegraphics[width=0.6\columnwidth]{HighRes/B_field_visualization}
%    \caption{Qualitative comparison of the $B$ and $E$-fields in the RF tracking region and the transverse drift region.\label{fig:b_field_vis}}
%\end{figure}

We propose to obtain this estimate using a novel experimental technique called \textit{cyclotron radiation emission spectroscopy} (CRES), which infers the kinetic energy of a charged particle from a measurement of the relativistic frequency shift of the cyclotron radiation emitted by the particle in a magnetic field \cite{cres_proposal}. This method was pioneered by the ongoing Project 8 neutrino mass experiment, which recently demonstrated an instrumental energy resolution of $1.66\pm0.19 \hspace{1mm} \mathrm{eV}$ \cite{PhysRevC.109.035503}. Thus, implementing a CRES-based electron tracker to access the parameters of electron motion at the ${\sim} 10\hspace{1mm}\mathrm{eV}$ energy resolution required by the PTOLEMY transverse drift filter is considered a feasible approach based on the current state of the art. This work describes a potential design and mechanism of a CRES-based electron tracker for the PTOLEMY experiment. While this study primarily focuses on the RF tracking of electrons within the uniform magnetic field region depicted in Fig. 2, the complete end-to-end transportation process from various target/source configurations to this region is an ongoing area of research. This work focuses on the electrons drifting with the endpoint energy at the center of the rectangular structure in between the pole faces of the magnet. Using electromagnetic simulations, we demonstrate the feasibility of extracting the dynamic parameters of electrons near the tritium endpoint from a ${\sim}\mathrm{1\hspace{1mm}fW}$ electron cyclotron signal, while adapting to the unique design requirements imposed by the other modules in the PTOLEMY experiment (which make existing implementations \cite{PhysRevC.109.035503}\cite{He6_cres} of CRES incompatible). Additionally, we propose a simple algorithm for matching a measured spectrum to its total kinetic energy and pitch angle.\\

The general approach of RF tracking of semi-relativistic electrons has the potential for wide application to tritium endpoint measurements.  The CRES signal has been shown to substantially suppress background events at the tritium endpoint  \cite{project8_withelectronics}.  Thus, combining a differential micro-calorimeter energy measurement with the time-of-flight coincidence of the non-destructive electron tracker measurement can be a powerful tool for suppressing backgrounds in differential measurements.\\

The rest of the paper is organized as follows.
In Section~\ref{sec:geometry}, we outline the proposed geometry of the electron tracker. We specify the orientation of the static electromagnetic fields and provide a visualization of the electron trajectories within the RF region. In Section~\ref{sec:expected_spectra}, we characterize key features of the cyclotron radiation frequency spectrum emitted by semi-relativistic electrons moving under the influence of the electromagnetic field configuration described in Section~\ref{sec:geometry}. We clarify the dependence of the expected frequency spectrum on both the total kinetic energy and the pitch angle with respect to the uniform magnetic field. 
In Section~\ref{sec:energy_drft}, we derive the characteristic time scales for cyclotron frequency drift due to radiative energy loss. In Section~\ref{sec:setup_timedomain}, we explain the setup of electromagnetic simulations in CST Studio Suite and briefly discuss the time domain results. In Section~\ref{sec:frequency}, we analyze the simulated cyclotron electron signals in the frequency domain. In Section~\ref{sec:matching_algorithm}, we introduce and demonstrate the performance of an algorithm that matches a particular frequency spectrum to its center frequency and bouncing period. We explain how this technique could be used in conjunction with a pitch-angle non-selective calibration source. Finally, in Section~\ref{sec:noisefloor_driftspeed}, we estimate the necessary signal duration to obtain a sufficient signal-to-noise ratio~(SNR), assuming a noise floor based on the specification of a low-noise amplifier.

\section{Geometry of the Proposed RF Tracker and Typical Electron Trajectories \label{sec:geometry}}
\begin{figure}[htbp]
    \centering
    \includegraphics[width=.8\columnwidth]{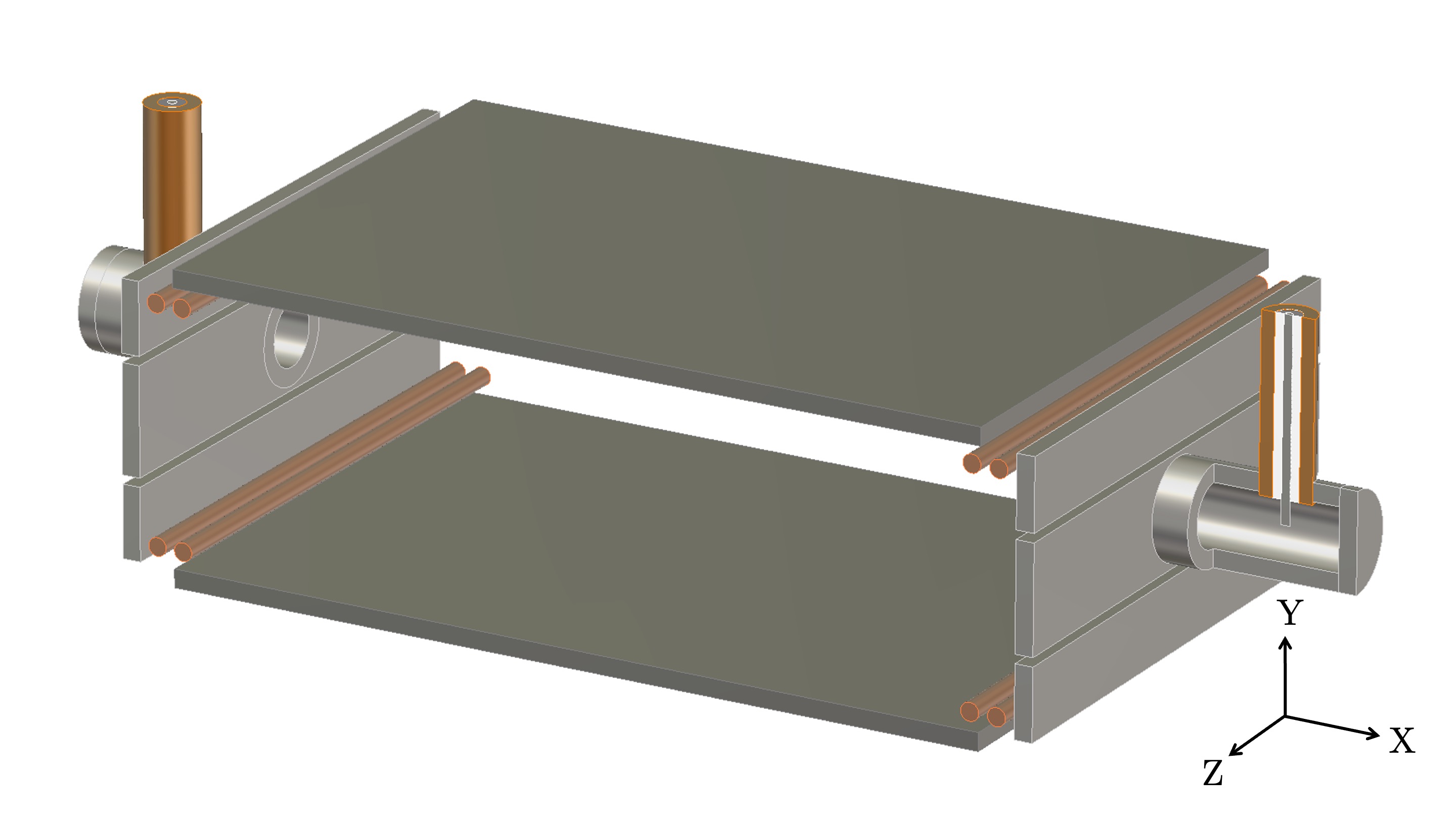}
    \caption{Geometry of the RF tracker. The antennae are coupled to cylindrical cavities on the lateral sides of the RF filter, as shown in the cut-away view on the right. The wires carry voltages to shape the electric potential to produce a more "bathtub"-like electric potential well. We present an image that is truncated in the $z$-direction to clarify the relative position of the cavity-coupled antennae; in reality, the RF tracking region extends longer in the $z$-direction. \label{fig:rf_tracker}}
\end{figure}

The basic structure of the RF tracker consists of four sets of boundary electrodes contiguous with those of the transverse drift filter, with radio-frequency antennas incorporated as an attachment to the existing geometry. In the specific design analyzed in this paper, a circular cavity is inserted into the bounce electrode with a probe pointing into the cavity to pick up an enhanced RF signal. This is represented schematically in Figure~\ref{fig:rf_tracker}. The $xy$-cross section extends $10~\mathrm{cm}$ in the $x$-direction, and $3~\mathrm{cm}$ in the $y$-direction. Like the rest of the electromagnetic filter, the RF tracker will be encased in a vacuum chamber (see Figure~\ref{fig:ptolemyconcept}).\\

The electromagnetic landscape in the RF tracking region is a superposition of three fields. First, a uniform $1~\mathrm{T}$ magnetic field is applied in the $x$-direction. This causes the electron to undergo circular motion in the plane normal to the magnetic field (i.e., the $yz$-plane). The frequency of the cyclotron motion is given by the relativistic formula
\begin{equation}
    f_{\mathrm{cyc.}} = \frac{eB}{2\pi\left(m + K/c^2\right)}~, \label{eq:f_cyc_constant}
\end{equation}

where $e$ is the elementary charge, $c$ is the speed of light, and $B$ is the magnitude of the uniform magnetic field in the $x$-direction. $m$ is the electron mass, while $K$ is the total electron kinetic energy. In the absence of axial motion, electrons emit radiation at a center frequency equal to the frequency of the cyclotron motion, with side bands peaked at integer multiples of the peak frequency. For electrons with kinetic energies in the vicinity of ${\sim} 18.6\hspace{1mm}\mathrm{keV}$, each $10\hspace{1mm}\mathrm{eV}$ difference in energy results in a $\approx 500\hspace{1mm}\mathrm{kHz}$ frequency shift. The power of the cyclotron radiation is given by the formula $P =  \left(\frac{2}{3}\right)\left(\frac{1}{4\pi \epsilon_0}\right) \frac{q^4 B^2}{m^2 c}\left(\gamma^2 - 1\right)\sin^2{\theta}$, where $\theta$ refers to the \textit{pitch angle}, the angle between the magnetic field vector $\vec{B} = B \hat{x}$ and the initial momentum vector of the $\beta$-decay electron. Electron energy loss due to RF emissions will be discussed further in Section~\ref{sec:setup_timedomain} and Appendix~\ref{appendix:B}. The cyclotron radius $\rho$ is given by $\rho = \frac{\gamma m v_{\perp}}{eB}$ \cite{jackson}. For $B=1~\mathrm{T}$, $K = 18600~\mathrm{eV}$, and electron momentum perpendicular to the magnetic field, we obtain $\rho \approx 0.5~\mathrm{mm}$; the variation of this value is negligible for pitch angles in our range of interest.\\

%%% A plot of the kinetic energy vs. center frequency. As pointed out by Andi, it may not be very informative since we already give an estimate above in terms of the rough spacing in center frequency for each 10 eV difference kinetic energy

%  \begin{figure}[htbp]
%    \centering
%\includegraphics[width=0.8\columnwidth]{HighRes/ke_freq_comparison.jpg}
%    \caption{Sample values of $f_{\mathrm{cyc.}}$ for electrons with kinetic energies near the tritium endpoint $E_0\approx 18577\hspace{1mm}\mathrm{eV}$ in a 1T $B$-field. Note that the $x$-axis labels are given as the \textit{difference} between $f_{\mathrm{cyc.}}$ and $27.0\hspace{1mm}\mathrm{GHz}$ in units of kHz.} \label{fig:freq_energy_dependence}
%\end{figure}

\begin{figure}[htbp]
    \centering
\includegraphics[width=0.9\linewidth]{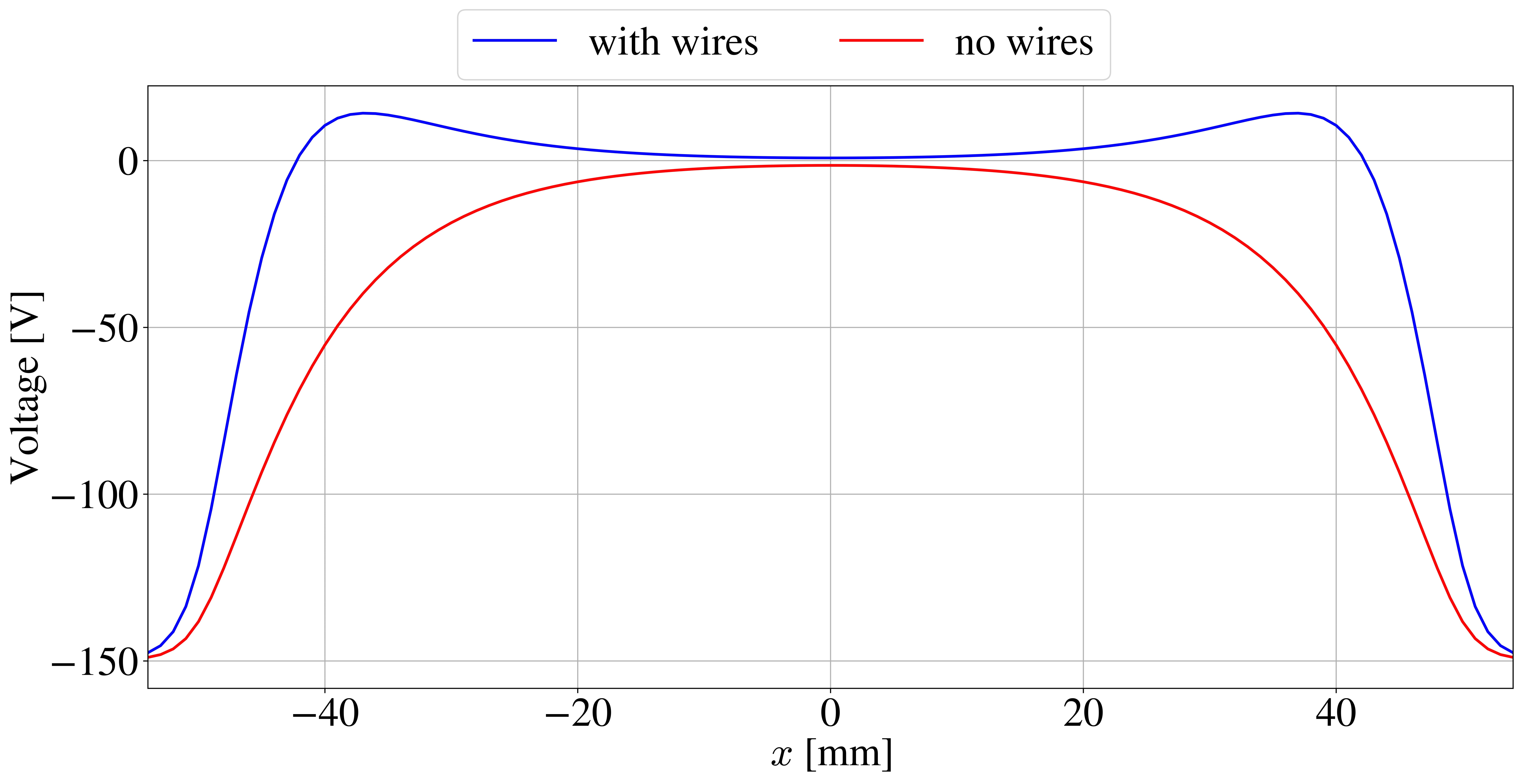}
    \caption{Electrostatic potential profiles caused by the $E_x$ field along $y=z=0$ for a geometry with (blue) and without (red) potential shaping wires. For $|x|>40\hspace{1mm}\mathrm{mm}$, the blue curve reaches the same potential values at larger values of $|x|$ than the red curve. This allows electrons with the same parallel momenta to approach closer to the side walls in the wire-shaped potential compared to the original potential.} \label{fig:potential_vis}
\end{figure}

\begin{figure}[htbp]
    \centering
\includegraphics[width=0.6\columnwidth]{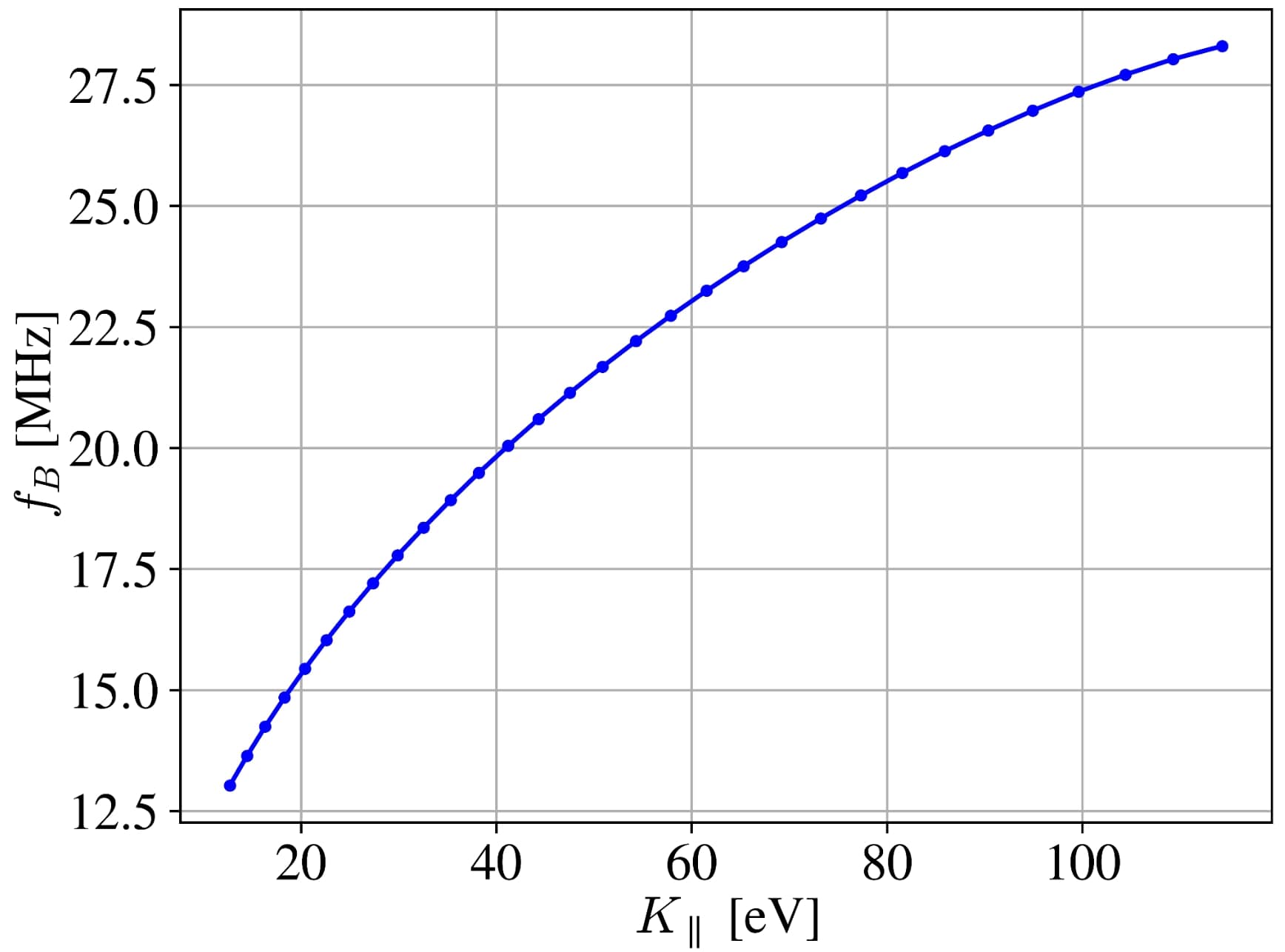}
    \caption{Frequency of the periodic guiding center motion along the $x$-axis as a function of electron $K_{\parallel}$, for the electrostatic potential profile shown in blue in Figure~\ref{fig:potential_vis}. \label{fig:bounce_freq}}
\end{figure}

The two side electrodes are set to equal, negative values, producing an electric potential well along the $x$-axis. This electrostatic potential well causes the guiding center of any electron with non-$90^{\circ}$ pitch angle to bounce back and forth in the $x$-direction. The frequency of this motion, $f_{B}$, depends on the  initial parallel momentum of the electron as well as the shape of the electrostatic potential. Wires are used to shape the electrostatic potential to be closer to a "bathtub" shape; this reduces the variation in the amplitude of the bouncing motion while increasing variation in $f_{B}$. A comparison of the electric potential profile along the line $y = z = 0$ for a geometry with (blue) and without (red) voltage shaping wires is visualized in Figure~\ref{fig:potential_vis}. 
The bouncing frequencies for different parallel kinetic energies calculated using particle trajectory simulations are presented in Figure~\ref{fig:bounce_freq}. For electrons with kinetic energies ${\sim}18.6~\mathrm{keV}$ and pitch angles $\gtrapprox 80^{\circ}$, the bouncing frequencies are of order ${\sim}\mathcal{O}(10)~\mathrm{MHz}$. 
The value of the side electrode voltages determines the maximum acceptable $K_{\parallel}$, which denotes the portion of the kinetic energy due to the momentum of the electron parallel to the magnetic field (i.e., $p_x$). Electrons with $K_{\parallel}$ greater than the absolute value of the bounce voltages will collide with the side walls, and cannot be transported through the filter. In this paper, we investigate ${\sim} 18.6\hspace{1mm}\mathrm{keV}$ electrons with pitch angles ${> 85^{\circ}}$, and thus the side electrodes were set to $-18600\cos^2{(85^{\circ})}\hspace{1mm}\mathrm{V}\approx -150\hspace{1mm}\mathrm{V}$. These voltages should be adjusted if a greater pitch angle acceptance range is desired. 

\begin{figure}[htbp]
    \centering
    \includegraphics[width=0.7\columnwidth]{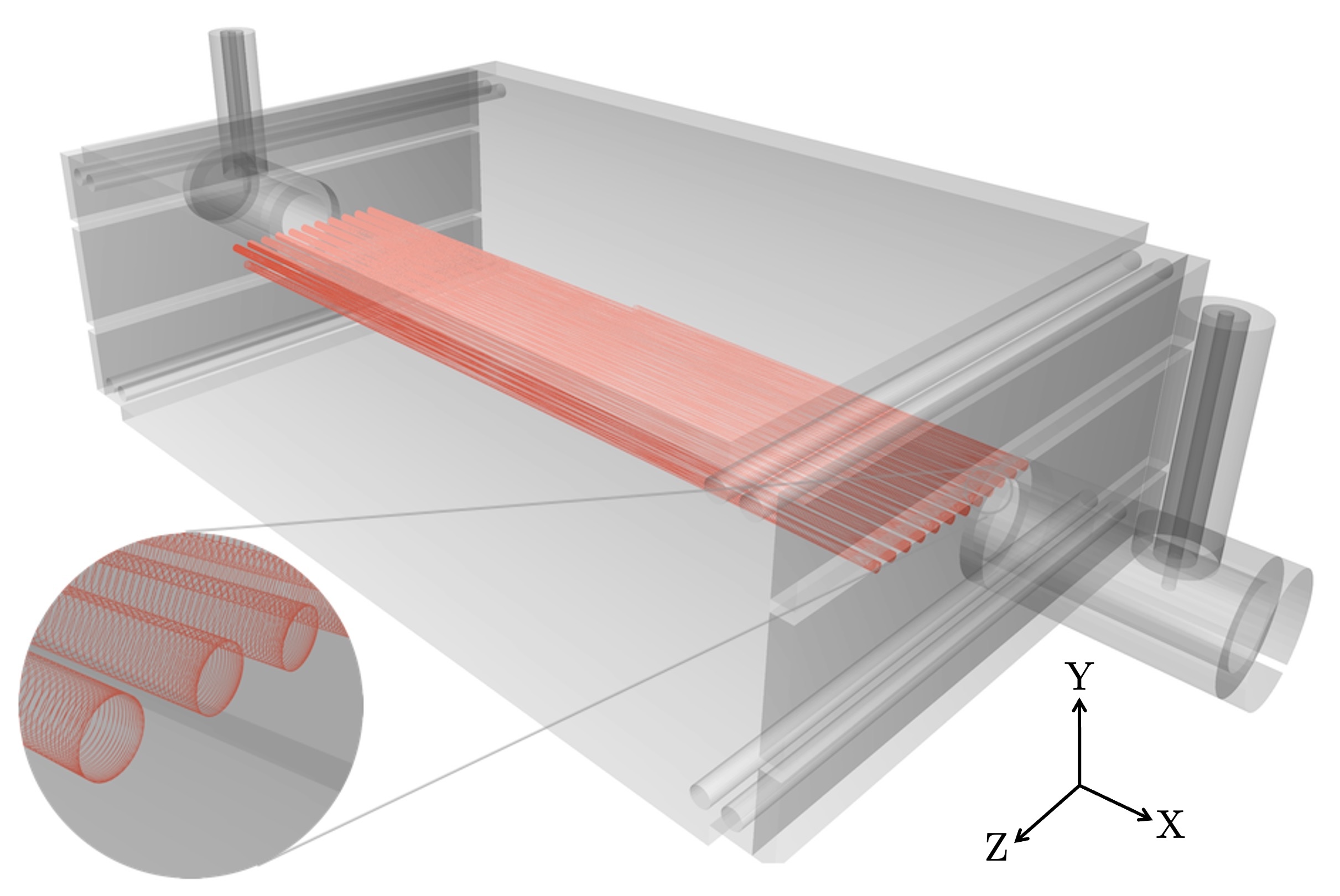}
    \caption{Sample trajectory of recoil electrons in the RF tracker presented in Figure~\ref{fig:rf_tracker}. Electrons undergo cyclotron motion in the $yz$-plane due to the magnetic field $\vec{B} = B \hat{x}$. The cyclotron guiding center bounces in the $x$-direction the electrostatic potential well due to the $E_x$ field produced by the electrodes. Finally, the electron undergoes slow drift in the $z$-direction due to $\vec{E}\times\vec{B}$-drift.   } \label{fig:example_traj}
\end{figure}

 Finally, a voltage difference is applied between the top and bottom plates in order to produce a uniform $E_y$ field. An electron undergoing cyclotron motion will undergo $\vec{E}\times \vec{B}$-drift of its guiding center with velocity $\vec{v} = \frac{\vec{E}\times \vec{B}}{|B|^2}$. In this geometry, $\vec{E}\times \vec{B} = E_yB_x \hat{z}= E_yB \hat{z}$; thus, the guiding center drifts in the $z$-direction. A sample trajectory is visualized in Figure~\ref{fig:example_traj}. The voltages on the electrodes positioned immediately above the cylindrical cavity are determined by the required drift velocity $v_z$ necessary to obtain a long enough signal to attain a suitable SNR.\\

Among numerous antenna configurations studied, we anticipate the potential of the cavity-coupled pin antenna. The geometry of the cylindrical cavity was determined to maximally enhance the resonant properties of the cavity at $\approx 27$~GHz while maintaining the aperture of the cavity as large as possible to maximize the solid angle of cyclotron radiation that can enter. We choose a waveguide radius of $a=3.5\hspace{1mm}\mathrm{mm}$, resulting in a vacuum cutoff frequency of 25~GHz for the $\mathrm{TE}_{11}$ mode (the fundamental mode for a circular waveguide) and a vacuum cutoff frequency of $\approx 32.8\hspace{1mm}\mathrm{GHz}$ for the higher mode ($\mathrm{TM}_{01}$). Thus, only the fundamental mode is allowed to propagate in the relevant frequency range in a $1~\mathrm{T}$ magnetic field, for either ${\sim} 18.6\hspace{1mm}\mathrm{keV}$ tritium endpoint electrons or $17.8\hspace{1mm}\mathrm{keV}$ $\mathrm{{}^{83m}Kr}$ conversion electrons planned as a monoenergetic calibration source. In simulation we observed a steep drop in efficiency for cavity radii $a > 3.8 \hspace{1mm}\mathrm{mm}$ and $a < 3.3 \hspace{1mm}\mathrm{mm}$, suggesting a narrow optimal range for the cavity dimensions.\\

The cylindrical cavity is coupled to a quarter wavelength pin antenna directly interfaced with a coaxial cable. We choose the length of the pin to be $\frac{\lambda}{4} = 2.78\hspace{1mm}\mathrm{mm}$, one-fourth of the vacuum wavelength at $27\hspace{1mm}\mathrm{GHz}$. The length $l$ of the cylindrical cavity was set to satisfy $l = \frac{n\lambda_g}{2}$, the condition for resonance at a given guide wavelength $\lambda_g$ \cite{pozar_mw}. We choose $n = 2$ so that $l = \lambda_g$, calculating $\lambda_g$ at $27\hspace{1mm}\mathrm{GHz}$. Since we will be operating at the $\mathrm{TE}_{11}$ mode and $a = 3.5 \hspace{1mm}\mathrm{mm}$, we obtain \cite{pozar_mw}
\begin{equation}
\lambda_g = \frac{2\pi}{\sqrt{k_{\mathrm{27\hspace{1mm}GHz}}^2-k_c^2}} = \frac{2\pi}{\sqrt{k_{\mathrm{27\hspace{1mm}GHz}}^2-\left(\frac{p'_{11}}{a}\right)^2}} \approx 30.1 \hspace{1mm}\mathrm{mm} ~,
\end{equation}
where $p_{nm}^{'}$ denotes the $n$-th zero of the derivative of the $m$-th Bessel function. The backshort distance $D$ (the distance between the antenna and the closed end of the cavity) was set to $6.0~\mathrm{mm}$ after scanning over values in the vicinity of the rule-of-thumb value, $D= \frac{1}{4}\lambda_g$.

\section{  Electron Cyclotron Radiation Spectra Received by an  On-Axis Observer\label{sec:expected_spectra}}
We derive several anticipated features of the radiation spectrum received by the electric field probe described in Section~\ref{sec:geometry}. Consider the electric field experienced by a stationary observer at a position $\mathbf{\vec{r_0}} = \left(x_0, 0,0\right)$ on-axis with the cyclotron orbit, at a time $t_{\mathrm{obs}} = \Delta(t) + t$. Note that we use $t$ to denote the \textit{retarded} time and  $\Delta(t) \equiv \frac{|\mathbf{r_0}-r(t)|}{c}$, where $r(t)$ is the position of the electron at $t$. The electric field produced by the gyrating electron has $x$, $y$, and $z$ components. However, only the field in the plane perpendicular to $x$ contains useful information. Thus, we disregard the $x$-component of the electric field throughout this analysis. Approximating the gyrating electron as a perfect source of circularly polarized radiation, the electric field at the point $\mathbf{r_0}$ is given by
\begin{equation}
\mathbf{E}(\mathbf{r_0}, t + \Delta(t)) =\operatorname{Re}\{ E_0\left(\hat{\mathbf{y}} \pm i{\hat{\mathbf{z}}}\right)\exp{\left[i2\pi \int_0^{t} f_\mathbf{r_0}(t')dt'\right]\}}~, 
\end{equation}
where $f_\mathbf{r_0}(t)$ is given by the expression
\begin{equation}
f_\mathbf{r_0}(t) = \frac{eB}{2\pi \left(m + K(t)/c^2\right)} \left[\sqrt{\frac{1+\beta(t)}{1-\beta(t)}}\right]~.
\label{eq:freq_function}\end{equation}
The function $K(t)$ describes the kinetic energy of the electron as a function of time, while $\beta(t)\equiv \frac{v_x(t)}{c}$, where $v_x(t)$ is the $x$-component of the electron velocity.  Precisely speaking, the quantity $f_{\mathbf{r_0}}(t)$ is the cyclotron frequency emitted at the retarded time $t$ multiplied by the relativistic longitudinal Doppler factor due to the guiding center velocity of the electron at the retarded time $t$. Note that this is not equivalent to the observed instantaneous frequency. Neglecting energy loss over short time scales $t \ll \mathcal{O}(1)~\mathrm{s}$\footnote{See Section~\ref{sec:energy_drft} for 
a calculation of energy drift due to RF emissions.},  $K(t)$ is periodic in $t$ with a fundamental period of $\frac{T_B}{2}$, and $\beta(t)$ is periodic in $t$ with fundamental period $T_B$. Thus, the function $f_{\mathbf{r_0}}(t)$ is periodic in $t$ with fundamental period $T_B$. We denote the fundamental frequency of $f_{\mathbf{r_0}}(t)$ as $f_B 
\equiv \frac{1}{T_B}$.\\

Since $f_{\mathbf{r_0}}(t)$ is periodic in $t$, the function $g(t)\equiv \int_{0}^{t}f_{\mathbf{r_0}}(t')dt'$ can be written as $g(t) = \overline{f}_{\mathbf{r_0}}t+h(t)$, where $\overline{f}_{\mathbf{r_0}}$ is a constant with units of frequency given by the expression
\begin{equation}
    \overline{f}_{\mathbf{r_0}} = \frac{\int_0^{T_B}f_{\mathbf{r_0}}(t') dt'}{T_B}~.\label{eq:freq_integral}
\end{equation}
The function $h(t)$ is some function that is periodic in $t$ with period $T_B$, while $\overline{f}_{\mathbf{r_0}}$ is the average value of $f_{\mathbf{r_0}}(t)$ over one period.\\

Dropping the $\operatorname{Re}\{ \}$ brackets for convenience, we can rewrite $s(t) \equiv \mathbf{E}(\mathbf{r_0},t + \Delta t)$ as
\begin{equation}
\begin{split}
s(t) &= E_0\left(\hat{\mathbf{y}} \pm i{\hat{\mathbf{z}}}\right)\exp{\left[i2\pi \left(\overline{f}_{\mathbf{r_0} }t + h(t)\right)\right]}\\ & = E_0\left(\hat{\mathbf{y}} \pm i{\hat{\mathbf{z}}}\right)\exp{\left[i2\pi \overline{f}_{\mathbf{r_0}}t\right]}\exp{\left[i 2\pi h(t)\right]}\\ 
&=
E_0\left(\hat{\mathbf{y}} \pm i{\hat{\mathbf{z}}}\right)\exp{\left[i2\pi \overline{f}_{\mathbf{r_0}}t\right]}\sum_{n=-\infty}^{\infty}a_n \exp{\left[i2\pi n f_B t\right]}
\label{eq:et_function}
\end{split}
\end{equation}
for some set of Fourier coefficients $a_n$. The Fourier transform of $s(t)$ is given by 
\begin{equation}
\begin{split}
 \hat{s}(\omega) &= \frac{1}{\sqrt{2\pi}} \int_{-\infty}^{\infty} dt e^{-i\omega t}\sum_{n=-\infty}^{\infty} a_n \exp{i2\pi\left[\left(n f_B + \overline{f}_{\mathbf{r_0}}\right)t\right]}~,
 \\
 \hat{s}(\omega) &\propto \sum_{n=-\infty}^{\infty} a_n \delta(\omega - 2\pi n f_B - 2\pi \overline{f}_{\mathbf{r_0}})~. \label{eq:fourier_product}
\end{split}
\end{equation}

This expression predicts that the frequency spectrum of the \textit{product} of the radiation emitted by the recoil electron and the time-dependent Doppler factor due to its parallel velocity exhibits a comb structure, consisting of sidebands spaced in integer multiples of the bouncing frequency $f_B$ away from a carrier frequency given by the average frequency of the radiation $\overline{f}_{\mathbf{r_0}}$ reaching the observer. We assert that this argument can be extended for the \textit{observed} spectrum. The Fourier transform with respect to $t_{\mathrm{obs}}$ takes the form
\begin{equation}
\propto \sum_{n=-\infty}^{\infty} b_n \delta(\omega - 2\pi n f_B - 2\pi \overline{f}_{\mathbf{r_0}}) ~, \label{eq:observed}
\end{equation}
such that the carrier frequency and sideband locations are identical to Equation~\ref{eq:fourier_product} but have magnitudes determined by another set of Fourier coefficients. This argument is summarized in Appendix~\ref{appendix:fourier_argument}.\\

From Equation~\ref{eq:observed}, it is evident that the carrier frequency of an electron cyclotron spectrum contains information about the total kinetic energy $K$ of the electron, while the spacing between the teeth, $f_B$, is dependent on $K_{\parallel}$. Both $\overline{f}_{\mathbf{r_0}}$ and $f_B$ also depend on the precise shape of the bouncing potential. This poses a problem for a non-pitch angle selective source, as slight variations in the pitch angle will shift the carrier frequency, effectively "smearing" its location. We verify and discuss this phenomenon using electromagnetic simulations in Section~\ref{sec:frequency}. Additionally, there is no general mechanism that would guarantee that the carrier frequency at $\overline{f}_{\mathbf{r_0}}$ would have the largest-magnitude Fourier coefficient, particularly after propagating through an antenna system and electronics chain with inevitably frequency-dependent gain. Thus, even correctly identifying the center frequency in a single electron spectrum can be potentially challenging. In Section~\ref{sec:matching_algorithm}, we propose a method for circumventing the problem of "carrier frequency smearing" during calibration with a source that is not pitch angle selective, combining our results from electromagnetic simulations in Section~\ref{sec:frequency}. Additionally, $\overline{f}_{\mathbf{r_0}}$, as well as the Fourier coefficients $b_n$, could, in principle, be calculated numerically for a specific electrostatic potential. We leave such studies for future work employing high-precision particle trajectory simulations. 

\section{Cyclotron Frequency Drift due to Radiative Energy Loss \label{sec:energy_drft}}

The cyclotron frequency given by Equation~\ref{eq:f_cyc_constant} depends on the kinetic energy of the electron; its value at the bottom of the electrostatic potential well $(x=0)$ will decrease over time due to radiative energy loss. We derive a relativistic expression for electron energy evolution as a function of time. We combine this result with Equation~\ref{eq:f_cyc_constant} to infer the time scale over which this energy drift significantly impacts the observed frequency spectrum. In Appendix~\ref{appendix:B}, we present an argument illustrating that the radiation due to linear acceleration at the turning points of the bouncing motion contributes negligibly to the total energy loss. Thus, this section focuses exclusively on electron energy loss due to cyclotron motion.\\

We start from the relativistic expression for the power radiated by a point charge with acceleration perpendicular to its velocity derived in \cite{jackson}, casting it into SI units:

\begin{equation}
P = \left(\frac{2}{3}\right)\left(\frac{1}{4\pi \epsilon_0}\right)\frac{e^2 c}{\rho^2}\beta_{\perp}^4\gamma^4 \label{eq:jackson_power}~.
\end{equation}
We use $\rho$ to denote the gyroradius of the cyclotron motion and $\beta_{\perp}$ to denote $\frac{v_{\perp}}{c}$. To aid numerical estimates, we choose to include all instances of $c$ explicitly. Substituting 
\begin{equation}
\rho = \frac{\gamma m v_{\perp}}{qB} = \frac{\gamma \beta_{\perp} m c}{qB}\label{eq:radius_equation}
\end{equation}
into Equation~\ref{eq:jackson_power}, we obtain
\begin{equation*}
P= \left(\frac{2}{3}\right)\left(\frac{1}{4\pi \epsilon_0}\right)\left(\frac{q^4 B^2 c}{\gamma^2 \beta_{\perp}^2 m^2 c^2}\right)\beta_{\perp}^4\gamma^4 =  \left(\frac{2}{3}\right)\left(\frac{1}{4\pi \epsilon_0}\right) \frac{q^4 B^2}{m^2 c}\beta_{\perp}^2\gamma^2 ~.
\end{equation*}
Using the relationship $\beta^2\gamma^2 = \left(\gamma^2 - 1\right)$, we obtain
\begin{equation}
P =  \left(\frac{2}{3}\right)\left(\frac{1}{4\pi \epsilon_0}\right) \frac{q^4 B^2}{m^2 c}\left(\gamma^2 - 1\right)\sin^2{\theta} ~.\label{eq:relativistic_power}
\end{equation}
Substituting $\gamma^2 = \left(E/mc^2\right)^2$ into Equation~\ref{eq:relativistic_power}, we obtain a differential equation for the total energy $E(t)$:
\begin{equation}
\frac{dE}{dt} = - \eta \left(E^2 - \left(mc^2\right)^2\right)~,\label{eq:diffeq}
\end{equation}
where we introduce $\eta$ to represent all multiplicative constants:
\begin{equation*}
\eta \equiv \left(\frac{2}{3}\right)\left(\frac{1}{4\pi \epsilon_0}\right)\frac{q^4 B^2}{m^4 c^5}\sin^2{\theta}~.
\end{equation*}
Equation~\ref{eq:diffeq} is a separable differential equation that can be solved using standard integration techniques. We obtain the solution
\begin{equation}
E(t) = mc^2 \left[\frac{\phi_0 \exp{\left(2\eta mc^2 t\right)} +1}{\phi_0 \exp{\left(2\eta mc^2 t\right)}-1}\right] ~,
\label{eq:rel_sol}
\end{equation}
where we use $\phi_0$ to denote a constant fixed by initial conditions:
\begin{equation*}
\phi_0 \equiv \frac{K_0 + 2mc^2}{K_0}~.
\end{equation*}
Here, $K_0$ is the initial kinetic energy of the charged particle. In the limit $t\rightarrow \infty$, Equation~\ref{eq:rel_sol} approaches $mc^2$, as expected.\\

In principle, Equation~\ref{eq:rel_sol} can be substituted into Equation~\ref{eq:f_cyc_constant} to solve for the time-dependent emitted cyclotron frequency at the bottom of the electrostatic potential well. For $t \ll \tau$, where $\tau = 1/(2mc^2\eta)$ is the characteristic time scale of Equation~\ref{eq:rel_sol}, Equation~\ref{eq:rel_sol} is quasilinear near $t=0$. For an electron in a 1~T magnetic field, we obtain $\tau\approx 2.58$~s. Thus, for $t\ll\tau$, we obtain
\begin{equation}
E(t)\approx \left(K_0 + mc^2\right) + Gt, \hspace{2mm} t\ll \tau~, \label{eq:linear_energy}
\end{equation}
where
\begin{equation}G \equiv \left(\frac{dE}{dt}\biggr\rvert_{t=0}\right) = P(0)=
-\frac{4\phi_0 \eta m^2c^4}{\left(\phi_0-1\right)^2} \approx -7.342 \hspace{1mm}\mathrm{meV}/\mathrm{\mu s}~, \label{eq:perp_eloss}
\end{equation}
where in the last term, we assume the specific values $B=1$~T, $K_0 = 18600\hspace{1mm}\mathrm{eV}$, and  $\theta = 90^{\circ}$ to make an estimate of the loss rate for $t\ll 2$~s in the relevant region of the parameter phase space.\\

For short time scales, Equation~\ref{eq:linear_energy} can be substituted into~\ref{eq:f_cyc_constant}, allowing us to obtain
\begin{equation}
f_{\mathrm{cyc.}}(t) \approx \frac{eBc^2}{2\pi \left(mc^2 + K_0 + Gt\right)},\quad t\ll\tau~.\label{eq:fcyc_changing}
\end{equation}

The rate of frequency drift is therefore 
\begin{equation}
\frac{df_{\mathrm{cyc.}}}{dt} \approx -\left(\frac{eBc^2}{2\pi}\right)\frac{G}{\left(mc^2+K_0+Gt\right)^2},\quad  t \ll \tau ~,
\end{equation}
where we recall that $G < 0$ and the cyclotron frequency increases as the electron loses energy. 
\begin{figure}[htbp]
    \centering
\includegraphics[width=\linewidth]{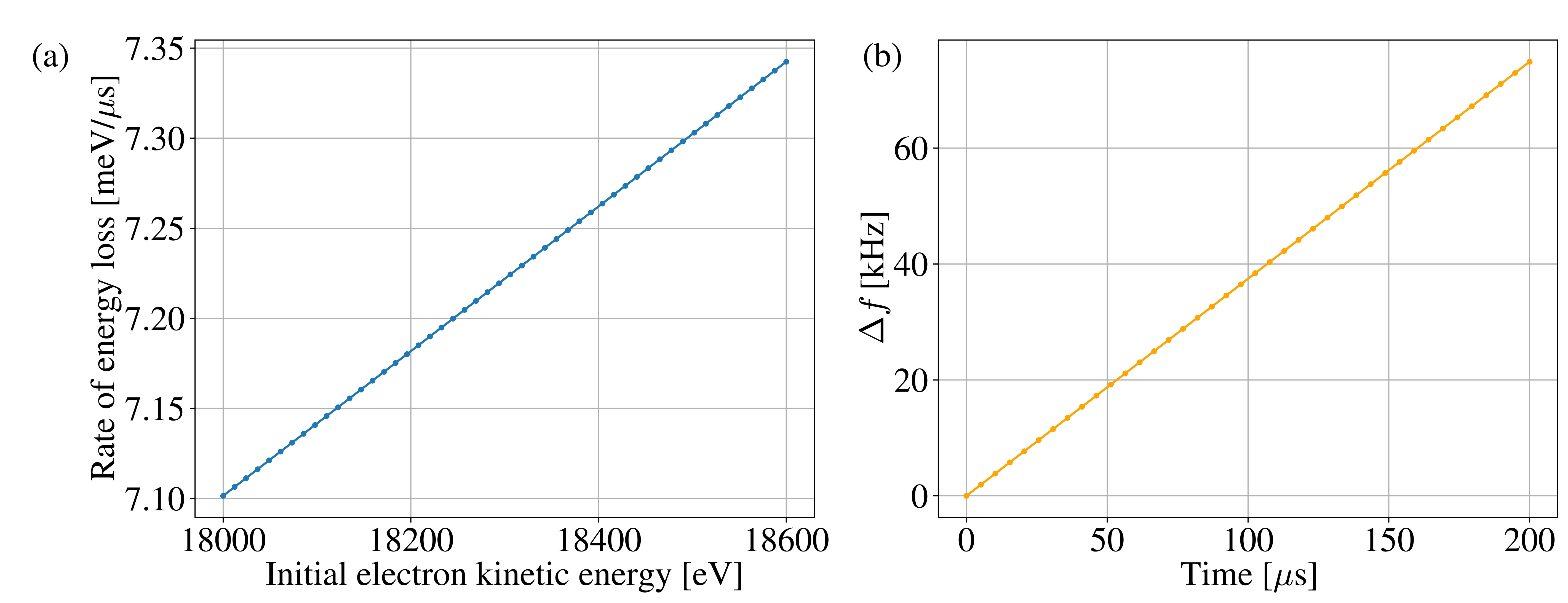}
    \caption{(a): Rate of energy loss for $t \ll \mathcal{O}(1) \hspace{1mm}\mathrm{s}$ for different initial kinetic energies near the tritium endpoint energy. (b): Deviation of the cyclotron frequency from the initial cyclotron frequency as a function of time due to radiative energy loss calculated using Equation~\ref{eq:fcyc_changing}, for an electron with $K_0 = 18600\mathrm{\hspace{1mm}eV}$ and $B=1$~T.
\label{fig:energy_losses}}
\end{figure}

In Figure~\ref{fig:energy_losses}(a), we present the rate of energy loss in the quasilinear regime for a range of initial electron kinetic energies. Variation in the energy loss rate between kinetic energies near the tritium endpoint is constrained to within roughly $\pm 1$ meV/$\mu$s. Meanwhile, in Figure~\ref{fig:energy_losses}(b), we plot the deviation of the cyclotron frequency from its initial value for a fixed initial kinetic energy $K_0 = 18600\mathrm{\hspace{1mm} eV}$. Estimating the slope of the curve in Figure~\ref{fig:energy_losses}(b), we ascertain that the rate of frequency drift is $\approx 375\hspace{1mm}\mathrm{Hz}/\mu$s.

\section{Simulation Setup and Time Domain Results \label{sec:setup_timedomain}}
To test the performance of the RF tracker proposed in Section~\ref{sec:geometry}, we model the apparatus in CST Studio Suite, a commercial electromagnetic solver software \cite{cst}. The metal surfaces were modeled as perfect electrical conductors, and the rest of the simulation domain is modeled as vacuum.\\

As discussed in previous sections, in any realistic implementation of the RF tracker, the $E_y$ field will be adjusted so that the drift of the guiding center in the $z$-direction is much slower than either the circular motion or the bouncing motion in the $x$-direction. Thus, to simplify our study, the voltages on the top and bottom of the tracker are set to $0\hspace{1mm}\mathrm{V}$ so that $E_y=0$. The simulation results presented in the following sections have zero drift in the $z$-direction. An analytic 1~T $B$-field was applied in the $x$-direction. We applied voltages of $-150\hspace{1mm}\mathrm{V}$ to the side electrodes and $+2\hspace{1mm}\mathrm{kV}$ to the potential shaping wires, producing the electrostatic potential given by the blue curve shown in Figure~\ref{fig:potential_vis}.\\

We choose a coordinate system so that the origin of the $yz$-grid coincides with the axis of the cylindrical cavity and $x=0$ corresponds to the center of the filter in the $x$-direction. The simulations were set up to begin such that the coordinates of the initial electron position are $x = z = 0$ and $y =  0.46 \mathrm{\hspace{1mm}mm} \approx \rho_c$, the radius of the cyclotron motion for an $18.6\mathrm{\hspace{1mm}keV}$ electron in a 1~T $B$-field (see Equation~\ref{eq:radius_equation}). Thus, in all simulations, the electron cyclotron motion occurred approximately on-axis with the cylindrical cavity, and the electron was released in the flat part of the potential where $\phi(x) \approx 0\hspace{1mm}\mathrm{V}$. The electron charge and mass were manually assigned to their most recent values published by the Particle Data Group \cite{pdg}. We chose to parameterize the kinematic properties of the electron in terms of total energy $K$ and pitch angle $\theta$. In each simulation, the electron was assigned a total energy $K$ and an initial velocity vector $\vec{v}(t=0)$. We chose to set $v_y(t=0)=0$ in all simulations. $v_x(t=0)$ and $v_z(t=0)$ were assigned to the values corresponding to a specific combination of $K$ and $\theta$.\\

We studied electrons with pitch angles ranging from $85.5^{\circ}$ to $88.5^{\circ}$ for several values of $K$ near the tritium endpoint.
For each combination of $K$ and $\theta$, we ran a $1.0\hspace{1mm}\mu\mathrm{s}$ simulation in the CST Particle-In-Cell (PIC) solver. This solver calculates particle dynamics within the predefined electromagnetic fields, the time-dependent electromagnetic fields produced by the moving particles, and the response of antennas in the time-dependent fields produced by these particle trajectories. Thus, trajectory calculation and antenna simulation are integrated in a single set-up. All simulations were terminated at $ 1\hspace{1mm}\mathrm{\mu s}$ due to numerical stability issues in the simulated antenna signal that were consistently observed after $\sim 1.5\hspace{1mm}\mu s$ of simulation.\\

\begin{figure}[htbp]
    \centering
    \includegraphics[width=0.9\linewidth]{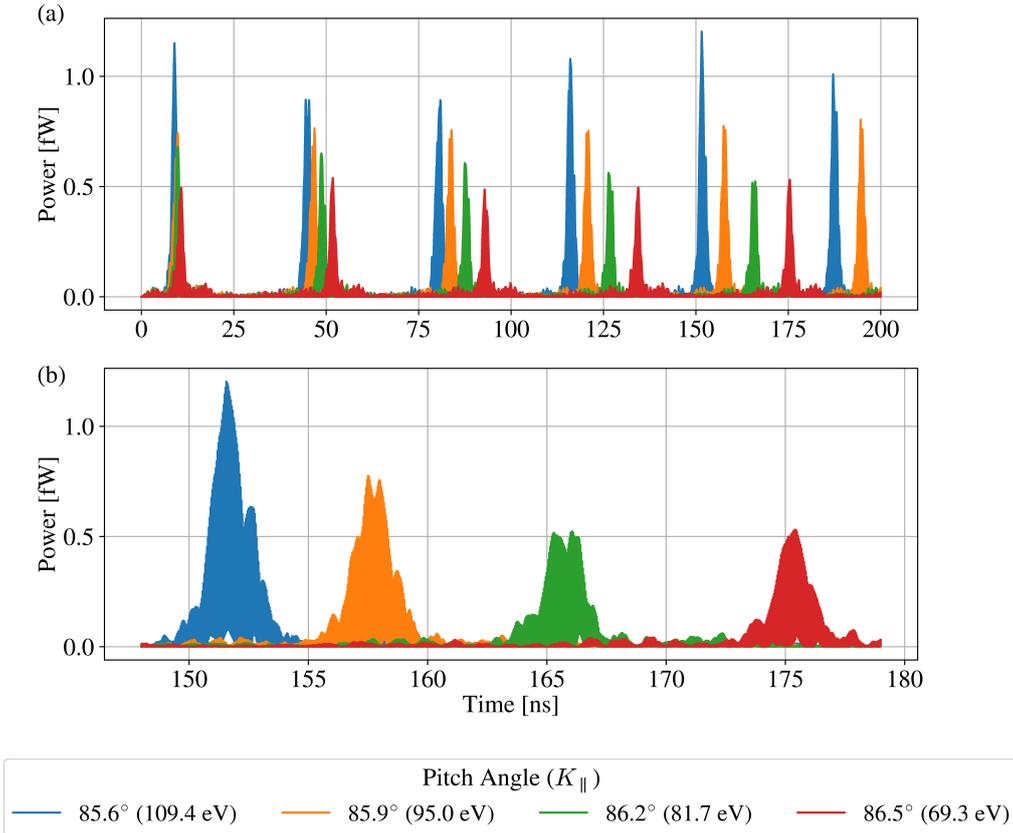}
    \caption{ (a): Cavity-coupled pin antenna response to electrons released into the simulation domain with $K =18590~\mathrm{eV}$ and various pitch angles ($K_{\parallel}$). The period of the antenna signal is clearly observed to vary depending on the value of $K_{\parallel}$. (b): Time signals from (a), plotted over a narrower time window to show the pulse width.
    \label{fig:time_domain_signal}}
\end{figure}

In Figure~\ref{fig:time_domain_signal}, we plot the power of the time domain signal as a function of time for fixed $K = \mathrm{18590.0\hspace{1mm}keV}$ and various pitch angles. The $K_{\parallel}$ value corresponding to each pitch angle is given in parentheses. We observe clear repeated pulses of high signal power roughly separated in time by the bouncing period $T_B = \frac{1}{f_B}$. For visual clarity, we plot only the first $200~\mathrm{ns}$ of each simulation; these pulses were observed to repeat with the same bouncing period for the rest of the $1~\mu$s simulations. In Figure~\ref{fig:time_domain_signal}(b), we plot the signals first presented in Figure~\ref{fig:time_domain_signal}(a) over a narrower time range to show the width of the high-power pulses; the pulses have time widths of $2{\sim}4\hspace{1mm}\mathrm{ns}$. The peak signal power tends to decrease for higher pitch angles, therefore, lower $K_{\parallel}$. We suspect that this is the result of the rapid reduction of the solid angle covered by the cylindrical cavity aperture observed from the electron as a function of distance from the waveguide; this effect could potentially be mitigated by applying higher voltages to the potential-shaping wires.

\section{Simulation Results in the Frequency Domain\label{sec:frequency}}
\begin{figure}[htbp]
    \centering
    \includegraphics[width=\linewidth]{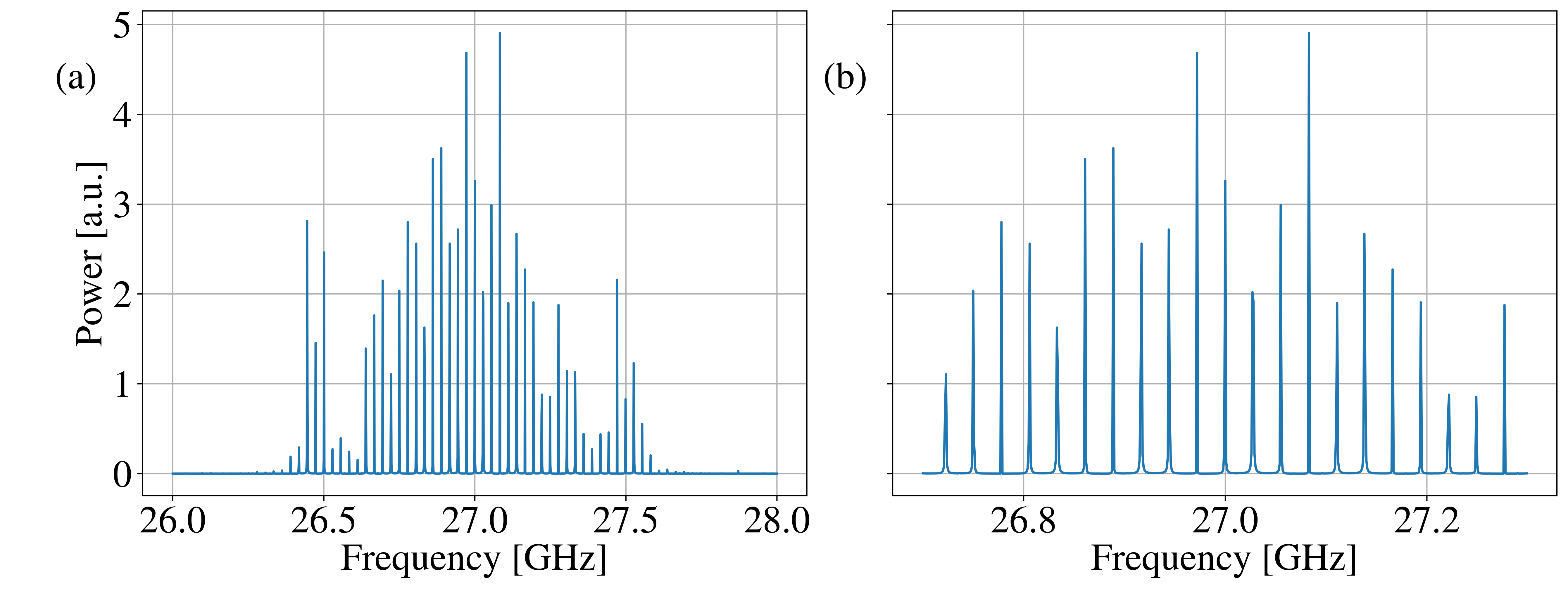}
    \caption{(a): Frequency spectrum received by an antenna due to radiation from a single electron. (b): Close-up version of the left, making it possible to estimate the width of the spacing. This is for $K = 18590~\mathrm{keV}$, pitch angle = $85.6^{\circ}$. 
\label{fig:single_electron}}
\end{figure}

We proceed to analyze the simulated antenna signals in the frequency domain. In Figure~\ref{fig:single_electron}, we present the FFT of the $1~\mathrm{\mu s}$ simulated signal due to a $K = \mathrm{18590~\mathrm{keV}}$, pitch-$85.6^{\circ}$ electron (blue curves in Figures~\ref{fig:time_domain_signal}(a) and (b)). We observe the comb structure expected from our discussion in Section~\ref{sec:expected_spectra}. The spacing between teeth is ${\sim} 28~\mathrm{MHz}$, consistent with the bounce frequencies calculated in Figure~\ref{fig:bounce_freq}. However, we observe that there does not exist a uniquely prominent peak that can identified as the carrier frequency; this introduces some difficulty in identifying the carrier frequency, henceforth denoted $f_c$. 

\begin{figure}[htbp]
    \centering
    \includegraphics[width=\linewidth]{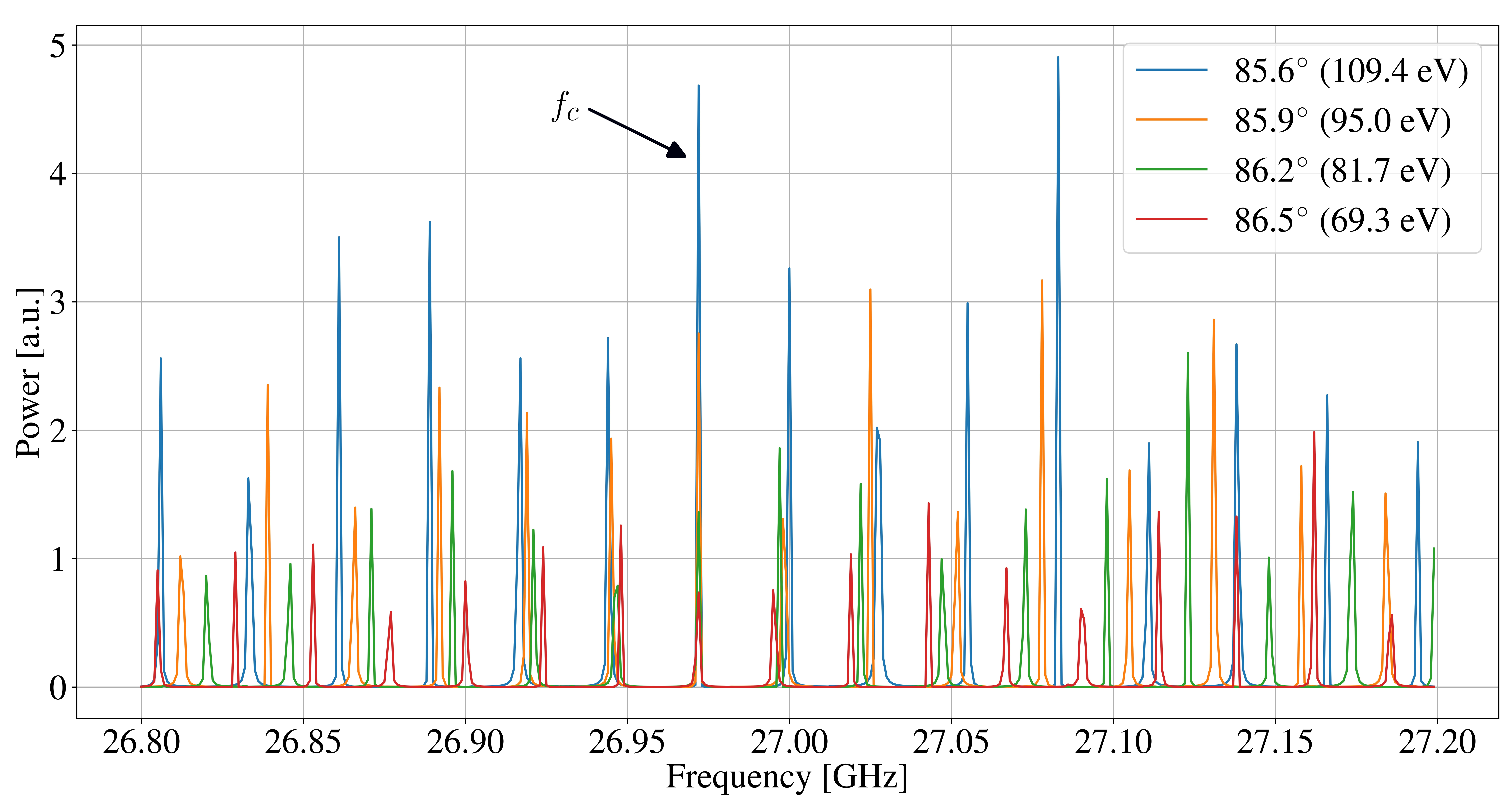}
    \caption{FFT of the four time domain signals plotted in the time domain in Figure~\ref{fig:time_domain_signal} produced by simulated electrons with $K = 18590\hspace{1mm}\mathrm{keV}$. The signals for all four pitch angles overlap at a common carrier frequency at $\approx 26.972\hspace{1mm}\mathrm{GHz}$. As expected from the discussion in Section~\ref{sec:expected_spectra}, the spectrum of each electron consists of sidebands offset by integer multiples of the corresponding bounce frequency $f_B$ from the carrier frequency, leading to decreasing overlap of the spectra in either direction from the carrier frequency.
\label{fig:multiple_electron_pitch}}
\end{figure}

In Figure~\ref{fig:multiple_electron_pitch}, we overlay the FFTs of the four $1~\mathrm{\mu s}$ time domain signals presented in Figure~\ref{fig:time_domain_signal}, over the frequency range $26.80{\sim}27.20\hspace{1mm}\mathrm{GHz}$ for visual clarity. We observe that for fixed $K = 18590\hspace{1mm}\mathrm{keV}$ and various pitch angles, all spectra align at $\approx 26.972~\mathrm{GHz}$. We identify this peak as the carrier frequency $f_c$ of all four frequency spectra.
The variation in the carrier frequency due to pitch angle is constrained to within $1\mathrm{\hspace{1mm}MHz}$, the frequency resolution of the FFT. The four spectra become less aligned in either direction away from this carrier frequency.\\   

\begin{figure}[htbp]
    \centering
    \includegraphics[width=0.8\linewidth]{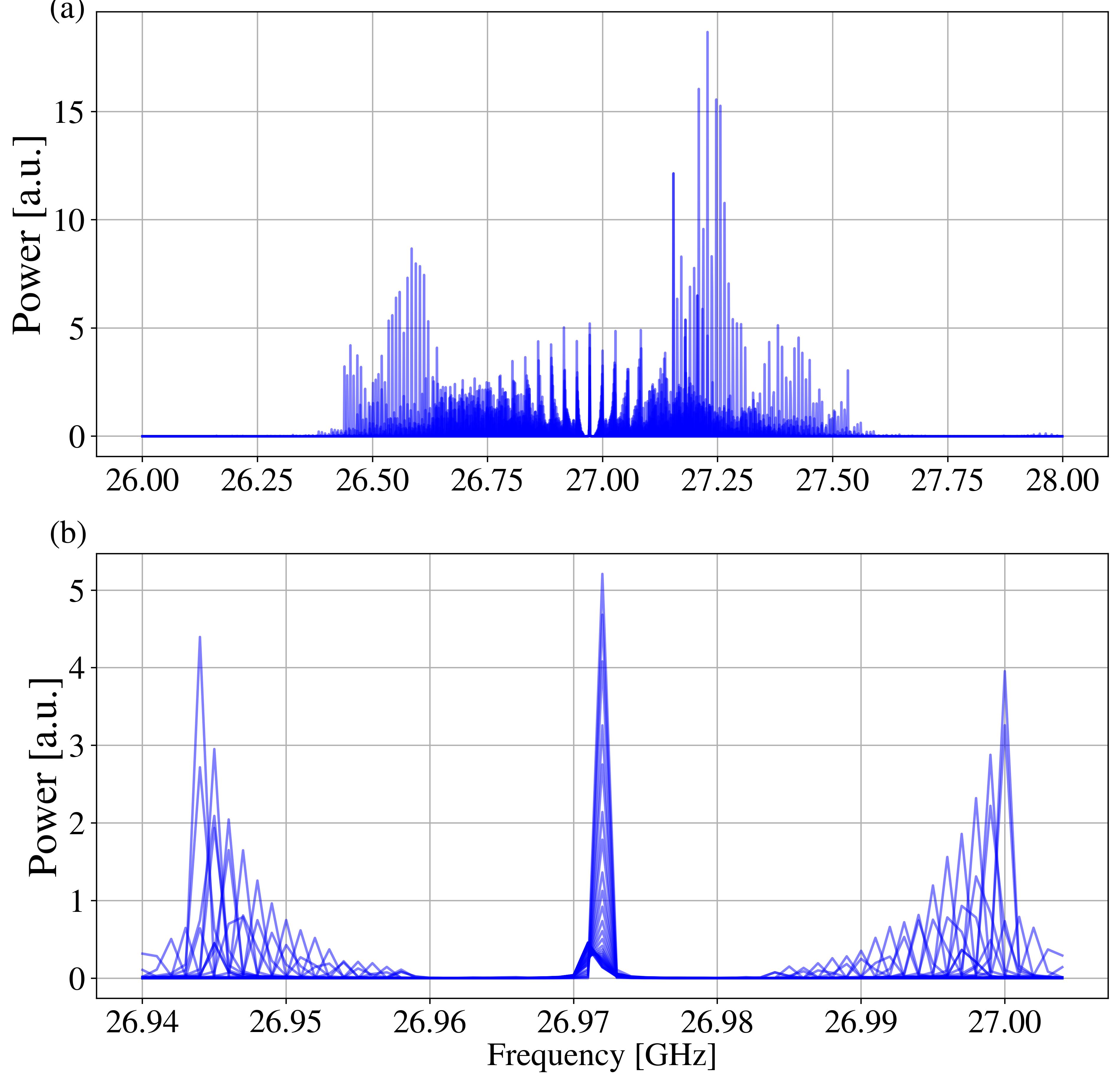}
    \caption{Overlayed FFTs of the time domain signals simulated for $K = 18590\hspace{1mm}\mathrm{keV}$ and all tested pitch angles between $85.5^{\circ}$ and $88.5^{\circ}$ ($0.1^{\circ}$ intervals). The carrier frequencies of the thirty spectra align at the same location, and can thus be distinguished from the sidebands.
\label{fig:all_pitches_ke18590}}
\end{figure}

\begin{figure}[htbp]
    \centering
    \includegraphics[width=1.1\linewidth]{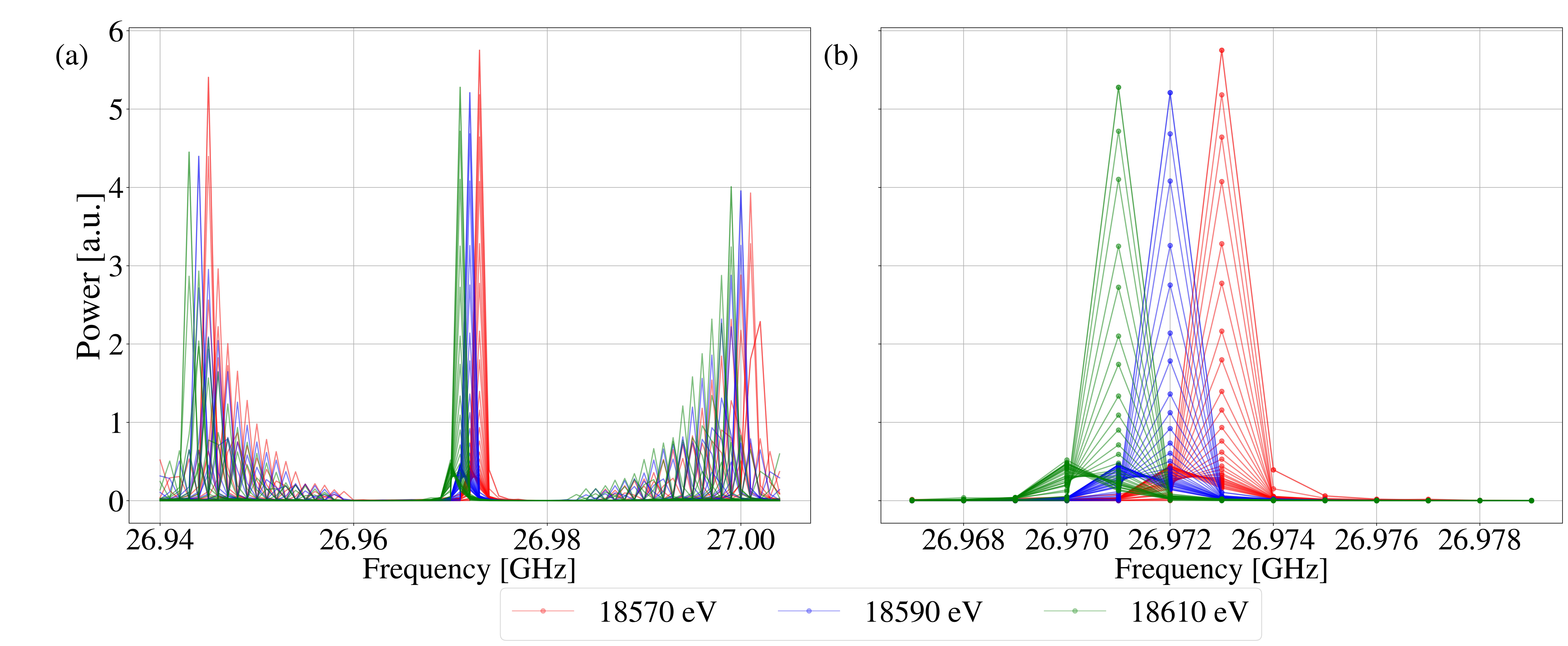}
    \caption{Overlayed FFTs of the time domain signals simulated for all tested pitch angles between $85.5^{\circ}$ and $88.5^{\circ}$ ($0.1^{\circ}$ intervals) for $K = 18570~\mathrm{eV, ~ 18590~\mathrm{eV},~\mathrm{and}~ 18610~\mathrm{eV}}$.
\label{fig:ke_comparison}}
\end{figure}

To investigate the variation in the carrier frequency due to Equation~\ref{eq:freq_integral} for a wider range of pitch angles with fixed $K$,  in Figure~\ref{fig:all_pitches_ke18590} we overlay thirty simulated spectra with pitch angles ranging from $85.5^{\circ}$ to $88.5^{\circ}$ in $0.1^{\circ}$ increments. All spectra appear to overlap at a clear location indicated in the plot. For higher pitch angles, the carrier frequency is shifted by 1 unit of frequency resolution; this can be seen more clearly in Figure~\ref{fig:ke_comparison}. This is consistent with the prediction made in Section~\ref{sec:expected_spectra} by Equation~\ref{eq:freq_integral}, anticipating a pitch-angle dependence on the carrier frequency $f_c$ for electrons with the same value of $K$. \\

One could picture that when calibrating the device using a monoenergetic but non-pitch angle selective electron source, the contributions due to many of these spectra would sum together, creating a prominent peak at ${\sim}f_{c}$ where the spectra overlap.  Although there would be some variation in the center tooth for electrons, its location within a $2{\sim}3\hspace{1mm}\mathrm{MHz}$ window can be clearly identified against a "background" of non-$f_c$ peaks. Based on these findings, in Section~\ref{sec:matching_algorithm}, we outline a calibration method assuming a monoenergetic but non-pitch angle selective calibration source, compatible with the matching algorithm that will be used to map a particular spectrum to a specific combination of $K$ and $K_{\parallel}$.\\ 
\\

Finally, in Figure~\ref{fig:ke_comparison}, we plot all simulated pitch angles for three kinetic energies offset by $20~\mathrm{eV}$: $K = 18570~\mathrm{eV, ~ 18590~\mathrm{eV},~\mathrm{and}~ 18610~\mathrm{eV}}$. The carrier frequencies for the three energies are each shifted by $\approx 1~\mathrm{MHz}$ relative to each other. This suggests that the carrier frequency of an observed spectrum can be feasibly used to map it to a specific electron kinetic energy,  provided that it can be reliably identified from the sidebands as the true carrier frequency.

\section{Spectral Matching Algorithm \label{sec:matching_algorithm}}
We found that for a fixed value of $K$, the carrier frequency $f_c$ is determined primarily by $K$ but also has a small pitch angle dependence. The spacing between sidebands also corresponds precisely to the physical bouncing frequency $f_B$, which depends only on $K_{\parallel}$. Since $f_B$ can be computed to high accuracy for a given $K_{\parallel}$ using only trajectory calculations, a precise measurement of $f_B$ is essentially equivalent to a measurement of $K_{\parallel}$. These observations motivate a spectral matching algorithm to carry out the required task of the PTOLEMY RF tracker, which is 1) to determine $K_{\parallel}$ and 2) to identify whether the total energy $K$ of an electron is within a finite energy interval $\Delta K^{\star}$ centered at a specific predetermined energy $K^{\mathrm{\star}}$ (e.g., the tritium endpoint). Calibration of the RF tracker with a monoenergetic but pitch angle non-selective electron source will be used to determine the frequency interval within which the carrier frequencies of signals produced by electrons with fixed total $K$ will be contained. We use $2\sigma_0$ to denote the width of the frequency interval and $f_0$ to denote the center of this interval. The empirically determined values of $f_0$ and $\sigma_0$ will then be used to generate a set of frequency templates, which are essentially square waves in frequency space. We refer to each nonzero region of the template spectra as \textit{teeth}. All templates for a given $K^{\star}$ contain the same tooth with center $f_0$ and width $2\sigma_0$, but the separation between successive teeth in each comb template corresponds to a different value of $f_B$.\\
\begin{figure}[htbp]
    \centering
    \includegraphics[width=\linewidth]{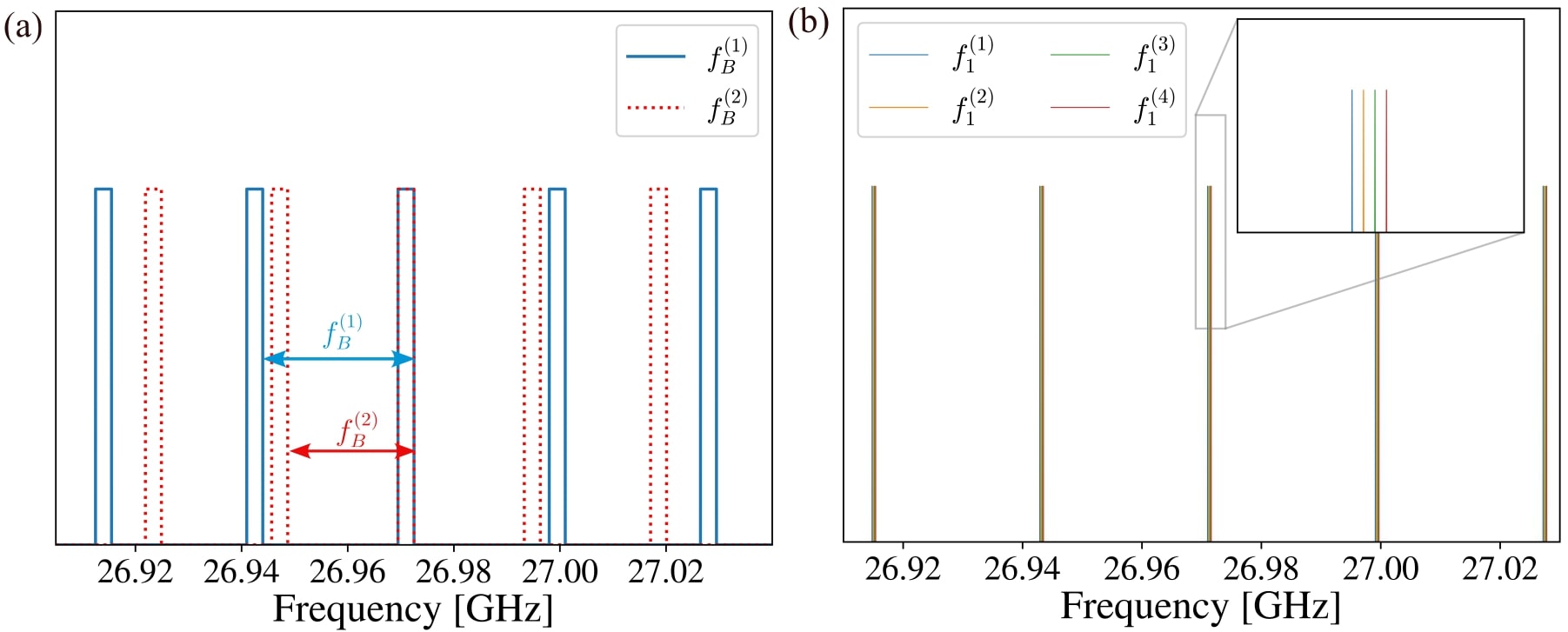}
    \caption{Schematic illustration of the comb template matching algorithm. (a): Wide-tooth comb templates for the same $K^{\star}$, characterized by the same $f_0$ and $\sigma_0$. The two templates test for two different values of $f_B$. (b): Narrow-tooth comb templates for a fixed $f_B$ and $f_0$ determined using the first step with the wide-tooth comb templates to be within the interval $[f_0-\sigma_0, f_0 + \sigma_0]$. In principle, the width of these teeth can be made as narrow as the instrumental frequency resolution.
\label{fig:template_schematic}}
\end{figure}

During experiment operation, a signal above the noise floor registered within the empirically determined target frequency interval $[f_0-\sigma_0, f_0 + \sigma_0]$ will trigger a search algorithm to find the frequency template with the best match to the measured spectrum. We illustrate this idea schematically in  Figure~\ref{fig:template_schematic}(a). There are many possible metrics by which the match between a frequency template and a measured spectrum could be assessed. One approach we found to perform well can be formulated as the following optimization problem:
\begin{equation}
    \argmin_{f_i}~\frac{\rvert A\lvert + \rvert B(f_i)\lvert - 2\lvert A \tilde{\cap} B(f_i)\rvert}{\rvert A \lvert + \rvert B(f_i) \lvert} ~,\label{eq:sim_metric}
\end{equation}
where we use $A$ to represent the measured spectrum and $B(f_i)$ to represent the frequency template generated with bouncing frequency $f_i$. $A = \{f_n\}_{n=1}^{N}$ represents the measured frequency spectrum as a set of $N$ frequencies consisting of a carrier frequency and its sidebands. Meanwhile, $B(f_i) = \{\left(f_m-\sigma_0, f_m+\sigma_0\right)\}_{m=1}^M$ represents the frequency template as a set of $M$ ordered pairs indexed by $m$. We use the notation $|A \tilde{\cap} B(f_i)|$ to denote an operation resembling that of the cardinality of the intersection of two sets: it counts the number of frequencies in $A$ which are within any of the $M$ frequency intervals specified by the ordered pairs in $B$. Thus, the numerator of the expression in Equation~\ref{eq:sim_metric} counts the number of "unique" elements in $A$ and $B(f_i)$. This number is normalized by the total number of elements in both sets. The frequency $f_i^{\star}$ that generated the template minimizing this metric is our estimate of the bouncing frequency.\\
\begin{figure}[htbp]
    \centering
    \includegraphics[width=0.8\linewidth]{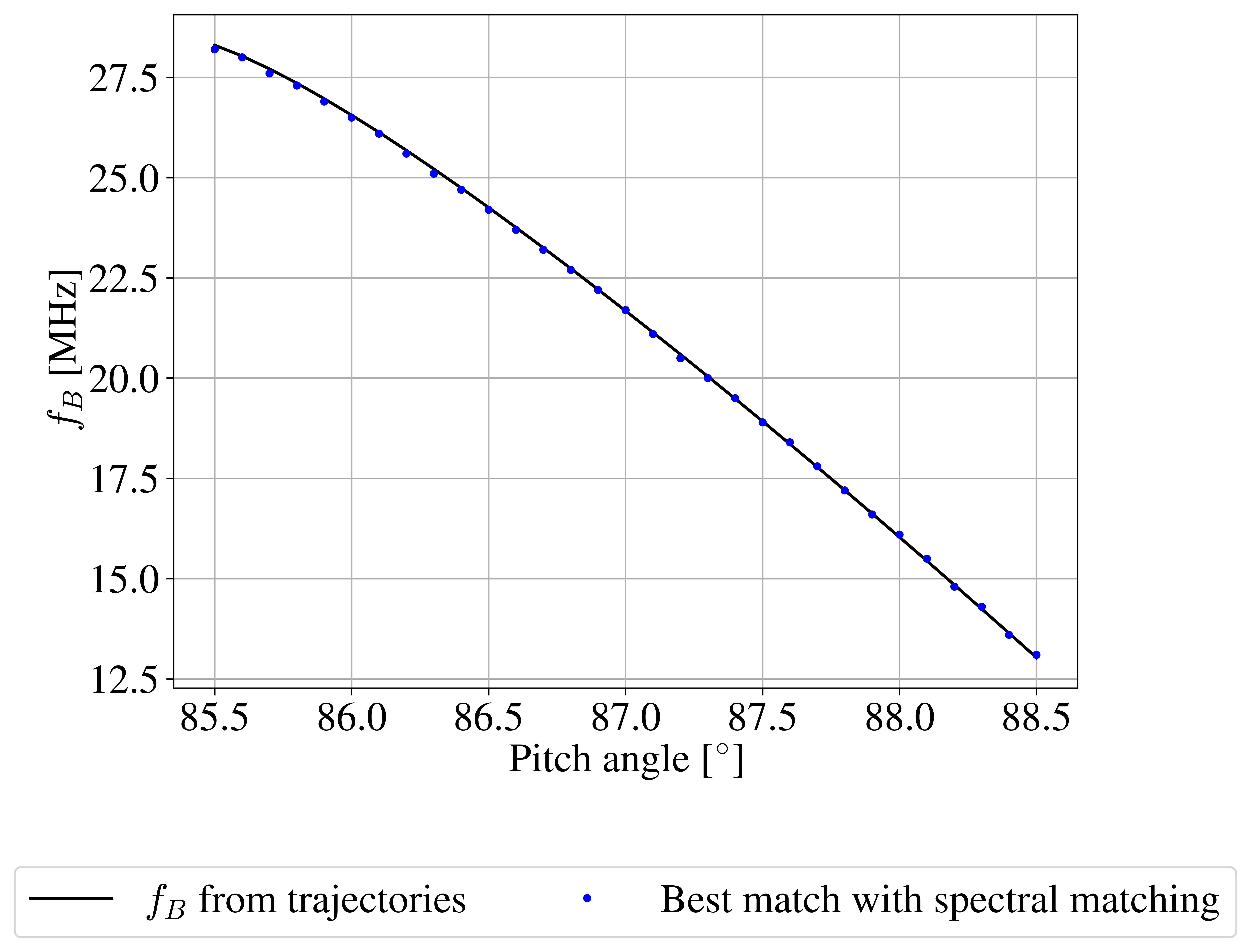}
    \caption{Result of the spectral matching algorithm for 30 pitch angles between $85.5^{\circ}$ and $88.5^{\circ}$ and $K = 18570~\mathrm{eV}$ (blue), compared to the bouncing frequencies calculated using electron trajectories. The mean squared residual between inferred vs. calculated $f_B$ is $\approx 58.55~\mathrm{kHz}$. 
\label{fig:spectral_matching}}
\end{figure}

In Figure~\ref{fig:spectral_matching}, we present our result from running this algorithm for the thirty simulated spectra with $K = 18570~\mathrm{eV}$. We let $f_0 = 26.973~\mathrm{GHz}$, treating the result in Figure~\ref{fig:ke_comparison} like a calibration measurement with a monoenergetic electron source.  We set $\sigma_0 = 2.5~\mathrm{MHz}$ and generated frequency templates spanning 0.5~GHz on either side of $f_0$. Templates were generated with $f_i$ ranging from 12 to 50~MHz in 100~kHz intervals. The simulated spectra were truncated to between 26.0~GHz and 28.0~GHz, assuming a receiver bandwidth of 2~GHz. The mean squared residual was $58.55~\mathrm{kHz}$. \\

Provided we have sufficient SNR, the procedure described above would correctly tag any electrons that have kinetic energies within the target energy interval and determine their bounce frequency with high accuracy. This estimate would then be used to adjust the parameters of the transverse drift filter and allow the electron to be guided to the calorimeter. On the other hand, this algorithm would produce false positives: any electrons emitting radiation at a carrier frequency outside of the target range $\Delta K^{\star}$, containing a sideband exactly coinciding with the target carrier frequency range, would also trigger the search algorithm. False positive electrons would not reach the calorimeter because the transverse drift filter voltages will not be tuned to that kinetic energy. These electrons would fail to be transported to the calorimeter, and would therefore not contribute to the measured spectrum. However, an excessive number of these false positives would be problematic if they are produced at a rate competitive with the speed at which the transverse drift filter can dynamically adjust the filter voltages and guide an electron to the calorimeter. Fortunately, we expect the number of false positives to be significantly limited due to the proposed target geometry, which will act as a pre-spectrometer that will cut off the energy spectrum to the highest ${\sim}1~\mathrm{keV}$. Thus, carrier frequencies will be limited to an effective range of ${\sim}50~\mathrm{MHz}$, and thus from Figure~\ref{fig:spectral_matching}, it is evident that the number of false positives for a given pitch-angle would be of order $\mathcal{O}(1)$. \\

 A narrower limit on $K$ could be, in principle, obtained by a more precise determination of $f_c$. After $f_B$ is estimated using the procedure above, one could repeat a similar analysis procedure with a set of templates with narrower teeth. This time, $f_B$ is fixed at $f_B^{\star}$ (the best match value of $f_B$), but the center of the new templates, which we denote $f_1$, is allowed to vary within the range $[f_0 - \sigma_0, f_0 + \sigma_0]$. The teeth width, $2\sigma_1$, of this second set of templates could be set as narrow as the frequency resolution of the measured spectrum. Then, the best match $f_c$ is identified using a search algorithm through comb templates, similar to $f_B$ described in the previous paragraph. The precise measurement of $f_c$, in combination with a sufficiently accurate model of $f_c$ as a function of $K$, $K_{\parallel}$, and tracker geometry, could be used to map a measured spectrum more precisely to a specific kinetic energy. The development and verification of such a model requires significantly more powerful computational resources than were available to us at the time of this work and is thus a task for future work.\\
 
We have thus demonstrated the application of a specific metric and algorithm adept at determining the bounce frequency of electrons. It is conceivable that alternative metrics and corresponding algorithms could accomplish this task with higher efficiency. We are optimistic that the collective advancements in rapid data processing techniques, pioneered by existing experiments in particle physics and further refined within the domain of gravitational wave astronomy, will significantly enhance the methodology proposed herein.

\section{Estimation of SNR with Respect to the Nyquist limit\label{sec:noisefloor_driftspeed}}

The electronics chain for processing our signals will consist of a cryogenic low-noise pre-amplifier module and a digital down-converter (DDC) module. The DDC will mix down the amplified signals in the frequency band centered at ${\sim}27\hspace{1mm}\mathrm{GHz}$ spanning a bandwidth of $1\hspace{1mm}\mathrm{GHz}$, down to a center frequency of ${\approx} 1~\mathrm{to}~1.5\hspace{1mm}\mathrm{GHz}$. The down-converted signal will be digitized at 5 gigasamples per second. The dominant noise contribution is expected to be Johnson-Nyquist noise from the cryogenic amplifier, which effectively sets a lower limit on the noise level. In this section, we estimate the SNR of the spectra obtained in Section~\ref{sec:frequency} with respect to the expected level of Johnson-Nyquist noise from the cryogenic amplifier.\\

We assume an equivalent noise temperature, $T_{n}$, of 15 K with the low-noise amplifier at approximately 5~K, as indicated by the LNF-LNC16\_28WB datasheet from Low Noise Factory~\cite{LNF_datasheet}. 
Translating into a noise power, we use $P_{\mathrm{noise}} = k_{B}T_{n}\delta \nu = 2.07\times 10^{-19}~\mathrm{W}$,
where $k_{B}$ is the Boltzmann constant, and $\delta \nu$ is 20~kHz, the frequency resolution given a measurement duration of 50~$\mu$s.\\

The noise floor is presumed to follow a Gaussian distribution. Under this assumption, the SNR will scale with the measurement duration at a fixed FFT resolution, as described by the Dicke radiometer equation:
\begin{equation*}
\mathrm{SNR} = \frac{P_{\mathrm{signal}}}{P_{\mathrm{noise}}} \sqrt{\tau *\delta\nu}~.
\end{equation*}
For a time-averaged signal power of 0.025~fW, the NR of a $50 ~ \mu s$ is calculated to be 6.04. In Section~\ref{sec:energy_drft}, we estimated that the frequency drift for $\beta$-decay electrons with $K{\sim}18.6~\mathrm{keV}$ is $\approx 375 ~\mathrm{Hz}/\mathrm{\mu s}$. A $50 ~ \mu s$ signal would result in a frequency drift $\Delta f_c {\sim} 20~\mathrm{kHz}$. With the frequency shift modeled, a $500 ~ \mu s$ signal would result in an SNR of 19 (12.8~dB).\\

The performance of the low-noise amplifier is crucial in achieving high SNR. Quantum-limited devices, such as a Josephson traveling-wave parametric amplifier~\cite{Qiu2023}, can significantly reduce noise levels. The SNR values discussed above are based on the average signal power from a single pin-cavity antenna. Employing orthomode transducers or multiple antennae can increase signal power, thus improving the SNR. These configurations can enhance detection capabilities and reduce noise, leading to more accurate measurements.

\section{Conclusion and Prospects}
\label{sec:conclusion}

We have demonstrated the feasibility of accessing the dynamic parameters of tritium $\beta$-decay electrons in the proposed RF tracker geometry. A significant improvement would involve transitioning to computational platforms capable of higher precision trajectory calculations. This is necessary for developing an accurate mapping from initial $K$ and $K_{\parallel}$ to a specific carrier frequency $f_c$ in a particular tracker geometry. This would enable a higher precision estimate of the electron's total kinetic energy, which could in theory surpass the calibration uncertainty associated with pitch angle variations.\\

Directions for further refinement would aim to extract additional information from cyclotron signals. For example, tracking the $y$-position of the drifting electrons would offer a pathway for suppressing surface background electrons. Meanwhile, deploying an array of antennae would open multiple avenues for augmenting the detection capabilities, in addition to improving the comprehensive fidelity of the experimental apparatus. An antenna array not only promises to substantially boost the SNR, but also to enrich our data with time-of-flight information.

\acknowledgments
Y.I. is a fellow of IRIS-HEP supported by the National Science Foundation under Cooperative Agreements OAC-1836650 and PHY-2323298.  This project was made possible through the support of grant (No. 62313) from the John Templeton Foundation.

\appendix
\section{Comb Structure of the Observed Spectrum\label{appendix:fourier_argument}}
In Section~\ref{sec:expected_spectra}, we showed that the product of the radiation emitted by a $\beta$-decay electron and the time-dependent Doppler factor due to its parallel velocity exhibits a comb structure in the frequency domain. Here, we extend this argument to the observed spectrum. We introduce a new function $\tau$ to express the emitted (retarded) time $t$ as a function of the observed time $t_{\mathrm{obs}}$, so that $t_{\mathrm{obs}} \equiv \tau(t)  = t + \frac{|\mathbf{r_0} - r(t)|}{c}$. Therefore, we can rewrite the second line of Equation~\ref{eq:et_function} as a function of $t_{\mathrm{obs}}$ by substituting $t = \tau^{-1}(t_{\mathrm{obs}})$:
\begin{equation} s(t_{\mathrm{obs}})= E_0\left(\hat{\mathbf{y}} \pm i{\hat{\mathbf{z}}}\right)\exp{\left[i2\pi \overline{f}_{\mathbf{r_0}}\tau^{-1}(t_{\mathrm{obs}})\right]}\exp{\left[i 2\pi h(\tau^{-1}(t_{\mathrm{obs}}))\right]}~. \label{eq:et_obs}
\end{equation}
Because $\frac{|\mathbf{r_0}-r(t)|}{c}$ is periodic in $t$ with period $T_B$, we can write $\tau^{-1}(t_{\mathrm{obs}}) = t_{\mathrm{obs}} + w(t_\mathrm{obs})$ where $w(t_\mathrm{obs})$ is some function periodic in $t_\mathrm{obs}$ with period $T_B$. This fact is apparent by comparison with the graph of the inverse of a function such as $f(x) = x+\cos{x}$. While  the inverse of this function cannot be expressed in terms of elementary functions, the graph of its inverse can be readily obtained by reflection about the line $y=x$; it is clear that the inverse function can also be written as the sum of the function $y=x$ and a second function periodic in $x$. \\

Recalling from Section~\ref{sec:frequency} that $h(t)$ is periodic in $t$ (by construction), $h(t)=h(\tau^{-1}(t_{\mathrm{obs}})) = h(t_{\mathrm{obs}} + w(t_{\mathrm{obs}}))$ is periodic in $t_{\mathrm{obs}}$ with period $T_B$. Let us define one final function of $t_{\mathrm{obs}}$: $h'(t_{\mathrm{obs}}) \equiv h(t_{\mathrm{obs}} + w(t_{\mathrm{obs}})) + w(t_{\mathrm{obs}})$. We realize that $h(t_{\mathrm{obs}} + w(t_{\mathrm{obs}}))$ is periodic in $t_{\mathrm{obs}}$ with period $T_B$, and thus $h'(t_{\mathrm{obs}})$ is periodic in $t_\mathrm{obs}$ with period $T_B$.  Thus,
\begin{equation}
\begin{split}
    s(t_{\mathrm{obs}}) 
    &= E_0\left(\hat{\mathbf{y}} \pm i{\hat{\mathbf{z}}}\right)\exp{\left[i2\pi \overline{f}_{\mathbf{r_0}}t_{\mathrm{obs}}\right]}\exp{\left[i2\pi w(t_{\mathrm{obs}})\right]}\exp{\left[i 2\pi (h(\tau^{-1}(t_{\mathrm{obs}})))\right]}\\
    &=E_0\left(\hat{\mathbf{y}} \pm i{\hat{\mathbf{z}}}\right)\exp{\left[i2\pi \overline{f}_{\mathbf{r_0}}t_{\mathrm{obs}}\right]}\exp{\left[i 2\pi (h'(t_{\mathrm{obs}}))\right]}~,
\end{split}
\end{equation}
 and the Fourier transform with respect to $t_{\mathrm{obs}}$ exhibits a comb structure similar to that in Equation~\ref{eq:fourier_product},
\begin{equation*}
\propto \sum_{n=-\infty}^{\infty} b_n \delta(\omega - 2\pi n f_B - 2\pi \overline{f}_{\mathbf{r_0}}) ~,
\end{equation*}
but for a  different set of coefficients $b_n$.
\section{Energy Loss due to Linear Acceleration of the Guiding Center\label{appendix:B}}
In Section~\ref{sec:energy_drft}, we claimed that the energy loss due to linear acceleration of the guiding center is negligible compared to energy loss due to perpendicular acceleration about the guiding center. A short argument justifying this approximation is presented here. We begin with the expression for radiated power due to linear acceleration in  \cite{jackson}, casting it into SI units:
\begin{equation}
P_{\parallel} = \frac{2}{3}\left(\frac{1}{4\pi \epsilon_0}\right)\left(\frac{e^2}{m^2 c^3}\right) \left(\frac{dp}{dt}\right)^2 = \frac{e^2 a^2}{6\pi\epsilon_0 c^3}
\end{equation}
 Consider a simplified "bathtub" electrostatic potential consisting of a flat region at the center and a harmonic potential near both walls:
\begin{equation}
V(x) = \begin{cases}
0, & |x| < x_0\\
\frac{1}{2} k(|x|-x_0)^2, & |x| > x_0
\end{cases} ~ .
\end{equation}
The expression $2x_0$ denotes the length of the flat portion of the potential. We let $2L$ denote the length of filter in the $x$-direction so that $-L \leq x \leq L$. Then the total dissipated energy over one period of the electron's trajectory in $x$ is given by
\begin{equation}
\Delta E = \frac{e^2}{6\pi\epsilon_0 c^3}\int_0^\frac{2\pi}{\omega} x_1^2 \omega^4 \cos^2{(\omega t)} dt = \frac{x_1^2 \omega^3 e^2}{6\epsilon_0 c^3} = \frac{x_1^2 e^2}{6\epsilon_0 c^3}\left(\frac{k}{m}\right)^{\frac{3}{2}}~,
\label{eq:delta_E}
\end{equation}
where $x_1=x_{\mathrm{max}}-x_0$, where $x_{\mathrm{max}}$ is the positive turning point of the electron motion in $x$.\\

We can see that $k$ and $x_1$ depends on $V_0$, the value of the potential at $x= L$, and $K_{\parallel,0}$, the initial value of the electron's $K_{\parallel}$. They are solved to be
\begin{equation*} k = \frac{2V_0}{\left(L - x_0\right)^2}, \hspace{3 mm}
x_1 = \sqrt{\frac{K_{\parallel,0}}{V_0}} \left(L-x_0\right)~.
\end{equation*}

Thus, Equation~\ref{eq:delta_E} simplifies to
\begin{equation}
\Delta E_{\parallel} = \frac{e^2 \sqrt{2}\left(K_{\parallel,0}\right)V_{0}^{\frac{1}{2}}}{3\epsilon_0 c^3 (L-x_0)m^{\frac{3}{2}}}~.
\end{equation}

Substituting sample values $V_0 = 150~\mathrm{eV}$,  $K_{\parallel,0} = 100~\mathrm{eV}$, $L = 0.05~\mathrm{m}$, and $x_0=0.04~\mathrm{m}$ into the above expression, we estimate $\Delta E_{\parallel} \approx -3\times 10^{-12}~\mathrm{eV}$ per bouncing period. \\

In comparison, from Equation~\ref{eq:perp_eloss} and Figure~\ref{fig:energy_losses}, we observe that the rate of energy loss due to perpendicular acceleration is $\approx 7~\mathrm{meV/\mu s}$, while from Figure~\ref{fig:bounce_freq} it is apparent that the bouncing period for an electron with $K_{\parallel,0} = 100~\mathrm{eV}$ is $T_B = \frac{1}{f_B} \approx \frac{1}{27.5 \mathrm{MHz}} \approx 36~\mathrm{ns}$. The energy loss due to cyclotron motion over one bouncing period is then given by $\Delta E_{\perp} \approx - 3\times 10^{-4}~\mathrm{eV}$. Thus, we conclude that the contribution of $\Delta E_{\parallel}$ to the total energy loss is negligible compared to $\Delta E_{\perp}$.  
% Bibliography

%% [A] Recommended: using JHEP.bst file
\bibliographystyle{JHEP}
\bibliography{main.bib}

%% or
%% [B] Manual formatting (see below)
%% (i) We suggest to always provide author, title and journal data or doi:
%% in short all the informations that clearly identify a document.
%% (ii) please avoid comments such as "For a review'', "For some examples",
%% "and references therein" or move them in the text. In general, please leave only references in the bibliography and move all
%% accessory text in footnotes.
%% (iii) Also, please have only one work for each \bibitem.

% \begin{thebibliography}{99}

% \bibitem{a}
% Author,
% \emph{Title},
% \emph{J. Abbrev.} {\bf vol} (year) pg.

% \bibitem{b}
% Author,
% \emph{Title},
% arxiv:1234.5678.

% \bibitem{c}
% Author,
% \emph{Title},
% Publisher (year).

% \end{thebibliography}
\end{document}